\documentclass[%
 aip,
 amsmath,amssymb,
 reprint,%
]{revtex4-1}

\usepackage{graphicx}
\usepackage{dcolumn}
\usepackage{bm}

\usepackage[utf8]{inputenc}
\usepackage[T1]{fontenc}
\usepackage{mathptmx}
\usepackage{etoolbox}

\makeatletter
\def\@email#1#2{%
 \endgroup
 \patchcmd{\titleblock@produce}
  {\frontmatter@RRAPformat}
  {\frontmatter@RRAPformat{\produce@RRAP{*#1\href{mailto:#2}{#2}}}\frontmatter@RRAPformat}
  {}{}
}%
\makeatother
\begin{document}

\preprint{AIP/123-QED}

\title[Interacting phosphorus and boron/aluminum $\delta$-doped layers]{First principles band structure of interacting phosphorus and boron/aluminum $\delta$-doped layers in silicon}
\author{Quinn T. Campbell}
\email{qcampbe@sandia.gov}
\affiliation{Center for Computing Research, Sandia National Laboratories, Albuquerque NM, USA} 
\author{Andrew D. Baczewski}
\affiliation{Center for Computing Research, Sandia National Laboratories, Albuquerque NM, USA}
\author{Shashank Misra}
\affiliation{Sandia National Laboratories, Albuquerque NM, USA}
\author{Evan M. Anderson}
\affiliation{Sandia National Laboratories, Albuquerque NM, USA}


\begin{abstract}
Silicon can be heavily doped with phosphorus in a single atomic layer (a $\delta$ layer), significantly altering the electronic structure of the conduction bands within the material.
Recent progress has also made it possible to further dope silicon with acceptor-based $\delta$ layers using either boron or aluminum, making it feasible to create devices with interacting $\delta$ layers with opposite polarity. 
Using Density Functional Theory, we calculate the electronic structure of a phosphorus-based $\delta$ layer interacting with a boron or aluminum $\delta$ layer, varying the distances between the $\delta$ layers. 
At separations 1 nm and smaller, the dopant potentials overlap and largely cancel each other out, leading to an electronic structure closely mimicking intrinsic silicon. 
At separations greater than 1 nm, the two $\delta$ layers behave independently of one another, with an equivalent electronic structure to a p-n diode with an intrinsic layer taking the place of the depletion region. 
One mechanism for charge transfer between $\delta$ layers at larger distances could be tunneling, where we see a  tunneling probability exceeding what would be seen for a standard silicon 1.1 eV triangular barrier, indicating that the interaction between delta layers may enhance tunneling compared to a traditional junction.  
\end{abstract}

\maketitle

\section{Introduction}

Atomic precision advanced manufacturing (APAM)~\cite{Ward2020EDFA,roadmap} makes it possible to place dopant atoms in silicon with single-atom precision \cite{fuechsle2010spectroscopy,fuechsle2012single} and concentrations beyond the solid-solubility limit \cite{ruess2007realization} with small width, i.e. $\delta$ layers.
This fabrication technology has been investigated for use in applications ranging from quantum computing and analog simulation to more conventional digital electronics.\cite{gao2020RTOperation,halsey2022Robustness,wang2022natcomm,jones2023acsnano,roadmap,anderson2025integration,lu2021path,gao2021modeling}
While APAM was originally developed for placing $n$-type dopants (principally phosphorus, but arsenic too \cite{stock2020atomic}), recent progress has included a better understanding of surface chemistries involving acceptor-bearing precursors and even the fabrication of devices with $p$-type dopants.\cite{skeren2018nanotechnology,radue2021alcl3,campbell2021model,campbell2022hole,dwyer2021area,campbell2022reaction}
This raises the possibility of creating bipolar electronic devices entirely in silicon using both donor and acceptor based APAM processes on the same system.

There are, however, relatively few explorations of the electronic structure of silicon subject to both $n$- and $p$-type APAM doping. \cite{vskerevn2020bipolar,mendez2025exploring} 
For the creation of devices based on bipolar stacks of $\delta$ layers, it is important to understand how the electronic structure of these $\delta$ layers would interact, particularly at small separation distances. 
One intriguing possibility is that the induced conduction bands near the center of the Brillouin zone from the P $\delta$-layer will have their charge compensated by the induced valence bands from the B $\delta$-layer, moving the Fermi level into the band gap and creating a direct gap semiconductor within silicon. Conversely, the relevant defect potentials may interfere with one another, producing an electronic structure more closely resembling that of intrinsic silicon. This possibility strongly affects the understanding of tunneling in bipolar devices, where it is important to understand the conditions that enhance or suppress tunneling. 

Prior theoretical work on the electronic structure of $\delta$-layers used Density Functional Theory (DFT) and  primarily focused on  calculating the band structure of individual $\delta$ layers.
DFT has been used to predict the electronic structure of phosphorus $\delta$ layers in silicon,\cite{carter2009electronic,carter2011phosphorus,drumm2013ab1,drumm2013ab2} with essential features of these predictions subsequently confirmed experimentally.\cite{miwa2013direct,mazzola2014determining,mazzola2014disentangling,miwa2014valley,mazzola2018simultaneous,holt2020observation,katzenmeyer2020JMR,young2023suppression}
Together, these investigations showed that phosphorus $\delta$ layers introduce a metallic impurity band below the bulk conduction band edge.
A similar convergence between DFT calculation and experimental characterization was also recently observed for Sb $\delta$ layers. \cite{strand2025direct}
Recently, Campbell \textit{et al.} have analyzed DFT band structures for $\delta$ layers comprised of boron or aluminum.\cite{campbell2023electronic} 
They reported a similar, but inverted change in the band structure, in which an impurity band is formed above the bulk valence band edge and the Fermi level is located within these bands.
The success of DFT in describing the electronic structure of $\delta$-doped layers suggests that extending this methodology to include the interaction of donor and acceptor $\delta$-doped layers will provide a useful first approximation of the electronic structure of multiple $\delta$-doped layered structures. 

In this manuscript, we use DFT to predict the electronic structure and local density of states of interacting phosphorus and boron/aluminum $\delta$-doped layers in silicon, building upon previous work that has only looked at a single type of $\delta$-doped layer. 
We use the strongly constrained and appropriately normed (SCAN) exchange-correlation functional\cite{sun2015strongly} as a reasonable compromise between the computational cost of more accurate treatments of screened exchange (e.g., hybrid functionals) and the inaccuracy of more simplistic semilocal treatments, again providing novel results.
Our calculations predict that each layer creates a $\delta$ potential that hosts an impurity band, with the opposite charges compensating and mutually reducing the binding energies relative to the band edges.
The extent to which the binding energies of the impurity bands are reduced is a function of the distance between the $\delta$ layers.
At smaller separation distances $<$ 1 nm, the two $\delta$-layers largely compensate each other and the electronic structure most closely resembles that of intrinsic silicon, making a direct gap unlikely.
At distances $>$ 1 nm, the two layers become essentially independent.
At these larger distances, the $\delta$ layers essentially form the electronic structure of a p-n junction with an intrinsic layer of silicon comprising the depletion region.
We use the local potentials calculated within DFT to estimate the interlayer tunneling probability as a function of the separation between layers.
It is enhanced relative to the tunneling probability for a triangular barrier with a height of the intrinsic silicon band gap (1.1 eV), indicating that interactions between the $\delta$ layers do lead to outcomes that would not be achieved with traditional diodes. 

\section{Methods}
\label{sec:methods}

\begin{figure}
 \includegraphics[width=\columnwidth]{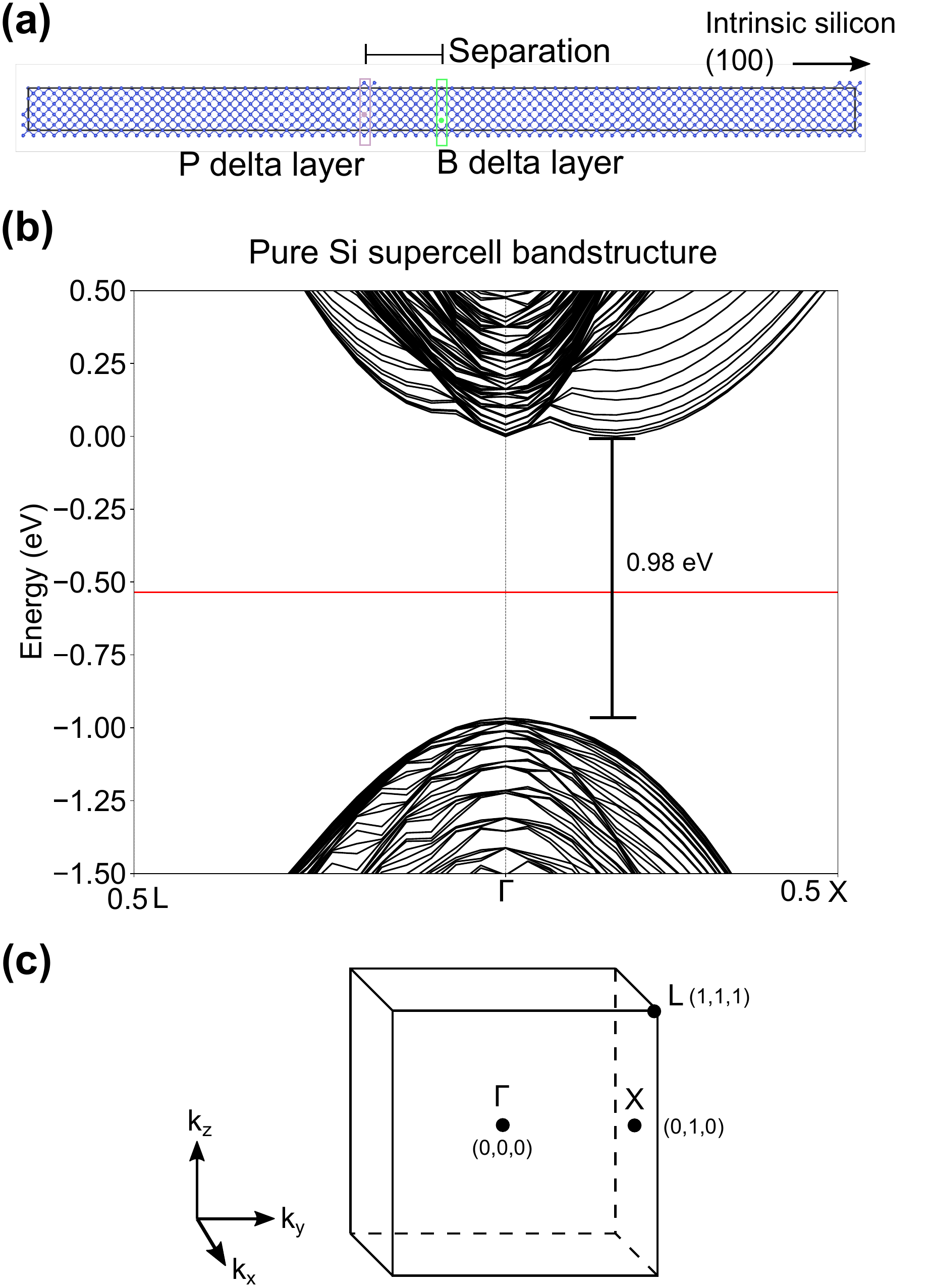}
\caption{(a) The supercell used throughout this manuscript, indicating the $P$ and $B$ $\delta$ layers and the distance separating them. The horizontal axis corresponds to the (100) direction. (b) The band structure of a pure silicon supercell. 
Because we are using a supercell, the more familiar Brillouin zone for a two-atom primitive cell is folded in on itself. The red line represents the Fermi level of the system. (c) The Brillouin zone for the supercell, labeling high-symmetry points with crystal reciprocal coordinates. The length of the box in the k$_x$ direction, associated with the (001) direction of the supercell is exaggerated for clarity. Given the significant length of the axis in real space, this dimension is essentially negligible in reciprocal space.}
\label{fig:schematic} 
\end{figure}

All of our calculations are based on 2$\times$1 Si(100) slabs (using the cubic unit cell as our base), as seen in Fig.~\ref{fig:schematic}a.
Two $\delta$-doped layers are placed within the silicon at 1/4 monolayer coverage.
Within these layers, all dopant atoms are placed substitutionally.
The rest of the supercell is pure silicon and we do not consider other features that might be expected in fabricated devices (e.g., other impurities, including the hydrogen or chlorine atoms that might be present due to the APAM doping process).
The supercell is 21.8 nm long, and we evaluate pairs of $\delta$-doped layers separated by as much as 10 nm. 
The periodic boundary conditions used in our calculations introduce an infinite array of images of each $\delta$-doped layer, with the closest being 21.8 nm away on either side of the (100) axis.
Prior analysis of single layers has found this distance to suffice for converging the effects of these images.\cite{carter2009electronic,campbell2023electronic}

All electronic structure calculations are done using the {\sc quantum espresso} package.\cite{giannozzi2009quantum}
We use the SCAN\cite{sun2015strongly} approximation to the exchange-correlation functional as implemented by Yao and Kanai.\cite{yao2017plane}
As shown in Fig.~\ref{fig:schematic}b, 
we predict a band gap of 0.98 eV for a pure silicon supercell. 
This is only slightly below the experimental value\cite{bludau1974temperature} of $\approx$1.1 eV, suggesting that estimates of other features of the band structure might achieve a similar degree of accuracy.
We use kinetic energy cutoffs of 680 eV and 2721 eV for the plane-wave basis sets used to describe the Kohn-Sham orbitals and charge density, respectively.
We use a 2$\times$2$\times$1 Monkhorst-Pack grid\cite{monkhorst1976special} to sample the Brillioun zone in our initial self-consistent calculation and then a 4$\times$4$\times$1 Monkhorst-Pack grid for non self-consistent calculations before band structures are calculated.
We reference all energies to the conduction band minimum (CBM) of the intrinsic silicon for any given structure.
Unless otherwise noted, the atomic positions are relaxed according to Born-Oppenheimer forces and considered converged below 0.5 eV/nm.
The cell size is kept fixed, mimicking the embedding of the $\delta$-doped layer within the larger silicon structure. 
Band structures and local density of states (LDOS) are then calculated for each structure. 

The use of a supercell leads to significant folding of the Brillouin zone of the usual bulk two atom primitive cell.
See the discussion by Drumm \textit{et al.}\cite{drumm2013ab1} for detailed interpretation of these band structures in $\delta$-doped layers. 
We illustrate the Brillouin zone used for this work in Fig.~\ref{fig:schematic}c.
Because of the significant Brillouin zone folding, we only plot out to half the direction of the special points, i.e. 0.5 X = (0.0,0.5,0.0) in crystal reciprocal coordinates.
It has previously been shown that the key characteristics of the band diagram can be captured near the $\Gamma$ point when examining DFT supercells for $\delta$ layer calculations.\cite{carter2013electronic}

In Appendix A, we show the band structures predicted for a single boron, aluminum, or phosphorus $\delta$-doped layer using the same supercell and exchange correlation functionals employed throughout this work. 
These results can be used to clearly isolate the impact of a single $\delta$-doped layer from the interaction of the multiple layers reported throughout this work.
In Appendix B, we show the doping potentials for boron and aluminum $\delta$-doped layers interacting with a phosphorus $\delta$-doped layer for each of the structures. 

\section{Results}
\label{sec:results}
\subsection{Band structure and LDOS of boron-phosphorus $\delta$-doped layers}
\begin{figure*}
 \includegraphics[width=\textwidth]{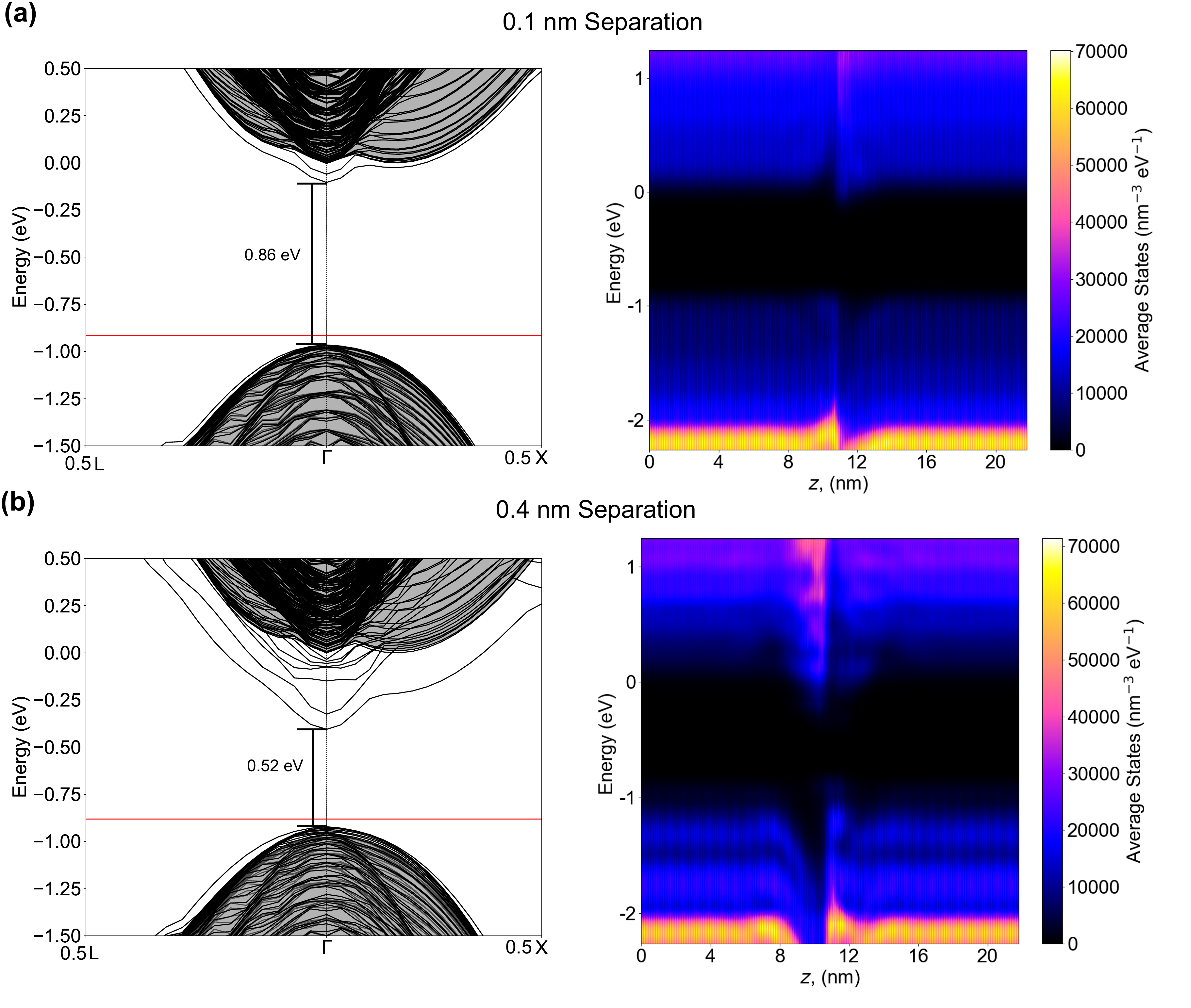}
\caption{The band structure and local density of states (LDOS) (i.e. the localized band structure) for boron and phosphorus $\delta$-doped layers separated by (a) 0.1 nm, and (b) 0.4 nm. As the separation distance increases, the $\delta$-layer potentials become more visible within the LDOS, decreasing the band gap of the overall structure. The red line represents the Fermi level of the system. 
}
\label{fig:bandstruc_low_sep} 
\end{figure*}

The band structure of a pair of phosphorus and boron $\delta$ layers separated by a single atomic layer ($0.124$ nm) is shown in Fig.~\ref{fig:bandstruc_low_sep}a.
At this range the layers compensate each other, leaving the material behaving largely as intrinsic silicon. 
The calculated distance from the top of the valence band to next band higher in energy is reduced slightly from 0.98 eV to 0.86 eV. It should be noted that the band diagram of two spatially separated $\delta$ layers can be somewhat misleading, as the z direction of the Brillouin zone is highly compressed. 
While the band structure appears to show a direct gap, we can see in the LDOS that the donor and acceptor $\delta$ potentials are spatially separated within the larger structure.
Nonetheless, simulations of the electronic transport in similarly spatially separated bipolar devices have shown that band gap narrowing likely plays a significant role in the resulting device behavior, \cite{mendez2025exploring} so we still report on the energetic separation between the donor and acceptor $\delta$ bands in figures within this work.
We avoid directly referring to this as a band gap since an electron would typically still need $\approx$1 eV energy to move directly from the valence band to the conduction band, according to the LDOS shown in Fig.~\ref{fig:bandstruc_low_sep}a. 
Given the strong charge screening at this short separation, the $\delta$ potentials more closely resemble two adjacent triangular potentials in the corresponding LDOS diagrams.
This separation distance is likely too small to be precisely achieved experimentally, but might be approximated through a sequence of dosing a silicon surface with one dopant precursor, annealing to incorporate it, and repeating the sequence with a different precursor without depositing silicon between dosing sequences.

At a separation of 0.4 nm, the $\delta$ potentials become much less suppressed, as seen in the band structure and LDOS in Fig.~\ref{fig:bandstruc_low_sep}b. 
This increase in the prominence of the $\delta$-doped layers can be attributed to the intervening silicon layers between the donors and acceptors screening the charge from each other.
The energetic distance from the boron $\delta$ layer peak to the phosphorus $\delta$ layer peak further narrows to 0.52 eV.
Furthermore, the $\delta$ potentials are now clearly visible in the LDOS in Fig.~\ref{fig:bandstruc_low_sep}b, showing a distinct protrusion for the phosphorus $\delta$ layer in the conduction band, and a lower magnitude, but still distinct protrusion in the valence band from the boron $\delta$ layer.

\begin{figure*}
 \includegraphics[width=\textwidth]{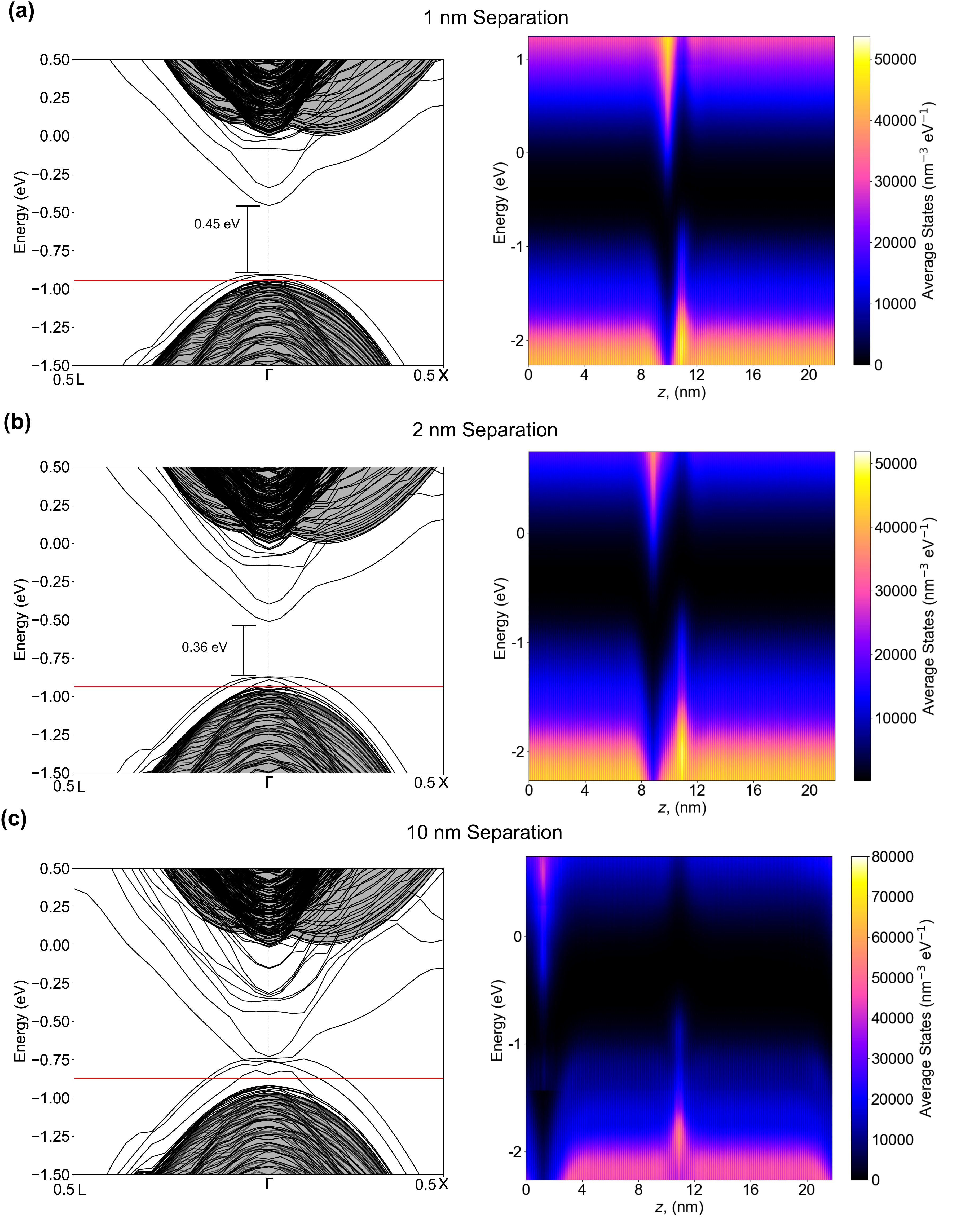}
\caption{The band structure and local density of states (LDOS) for boron and phosphorus $\delta$-doped layers separated by (a) 1 nm, (b) 2 nm, and (c) 10 nm. At these larger separation distances, the $\delta$-layer potentials become clearly distinct and, in the case of a 10 nm~separation, overlapping in energy. The red line represents the Fermi level of the system. 
}
\label{fig:bandstruc_high_sep} 
\end{figure*}

At separations of 1 and 2 nm, shown in Fig.~\ref{fig:bandstruc_high_sep}a and b, respectively, the different $\delta$ layers become clearly distinct in the resulting LDOS diagrams. 
At 1 nm, the induced VBM from the boron $\delta$-doped layer becomes clearly visible, shifting the Fermi level of the material within the valence bands. 
This is reflected in the energy difference from the boron $\delta$ layer peak to the phosphorus $\delta$ layer peak shifting to 0.45 eV at 1 nm, and clearly visible $\delta$ potentials in the LDOS in both the conduction and valence band.
This trend is continued at 2 nm, with an energy difference between the boron $\delta$ layer peak to the phosphorus $\delta$ layer peak of 0.36 eV and slightly less sharp $\delta$ potentials in the LDOS.
The two $\delta$-doped layers are still interacting and suppressing the expression of each other, but this suppression reduces with separation, thus explaining the gradually decreasing energy between the boron and phosphorus $\delta$ layer peaks. 

At a distance of 10 nm, however, the two $\delta$-doped layers are fully independent.
This leads to the induced valence and conduction bands from the $\delta$-doped layers overlapping in the middle of the band gap, as shown in Fig.~\ref{fig:bandstruc_high_sep}.
This is in line with what we would expect from an independent phosphorus $\delta$-doped layer with a magnitude $\approx$ 0.6 eV\cite{carter2009electronic} superimposed on an independent boron $\delta$-doped layer with a magnitude of $\approx$ 0.4 eV \cite{campbell2023electronic} away from the CBM and VBM, respectively.
It should be cautioned, however, that while the overall band structure for this separation looks like a metal, examining the LDOS shows that the valence and conduction band peaks are spatially confined to the $\delta$-layer positioning and thus unlikely to directly overlap.
Thus, this structure at 10 nm layer spacing is more akin to an ultra-short p-n junction, with a level of precision in dopant placement not attainable through traditional methods such as ion implantation. 

\begin{figure*}
 \includegraphics[width=\textwidth]{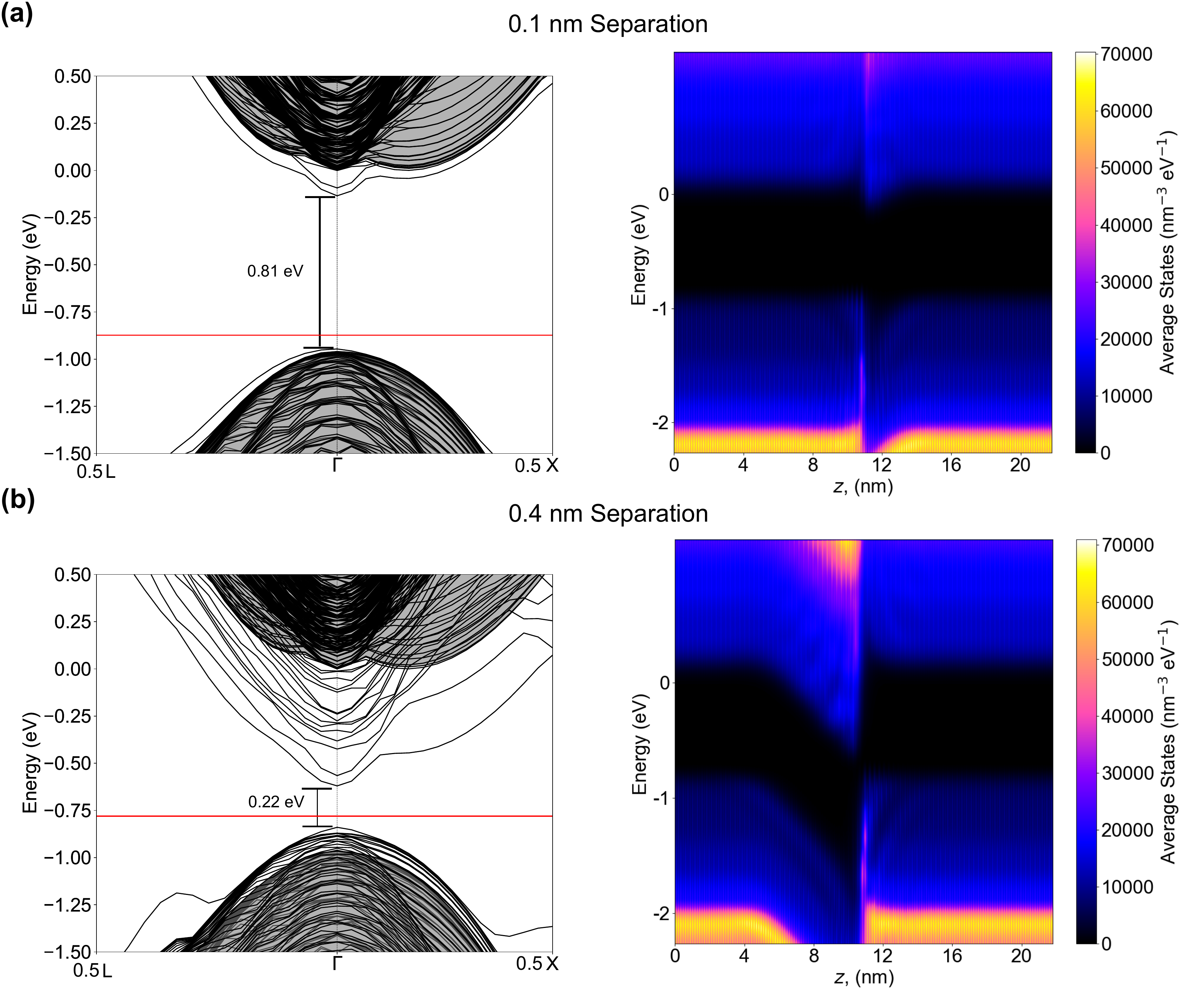}
\caption{The band structure and local density of states (LDOS) for aluminum and phosphorus $\delta$-doped layers separated by (a) 0.1 nm, and (b) 0.4 nm. The aluminum atoms have less suppression of the $\delta$ doped layer induced bands, resulting in overlap even at these lower separation distances. 
}
\label{fig:al-p-bandstruc-low} 
\end{figure*}

\begin{figure*}
 \includegraphics[width=\textwidth]{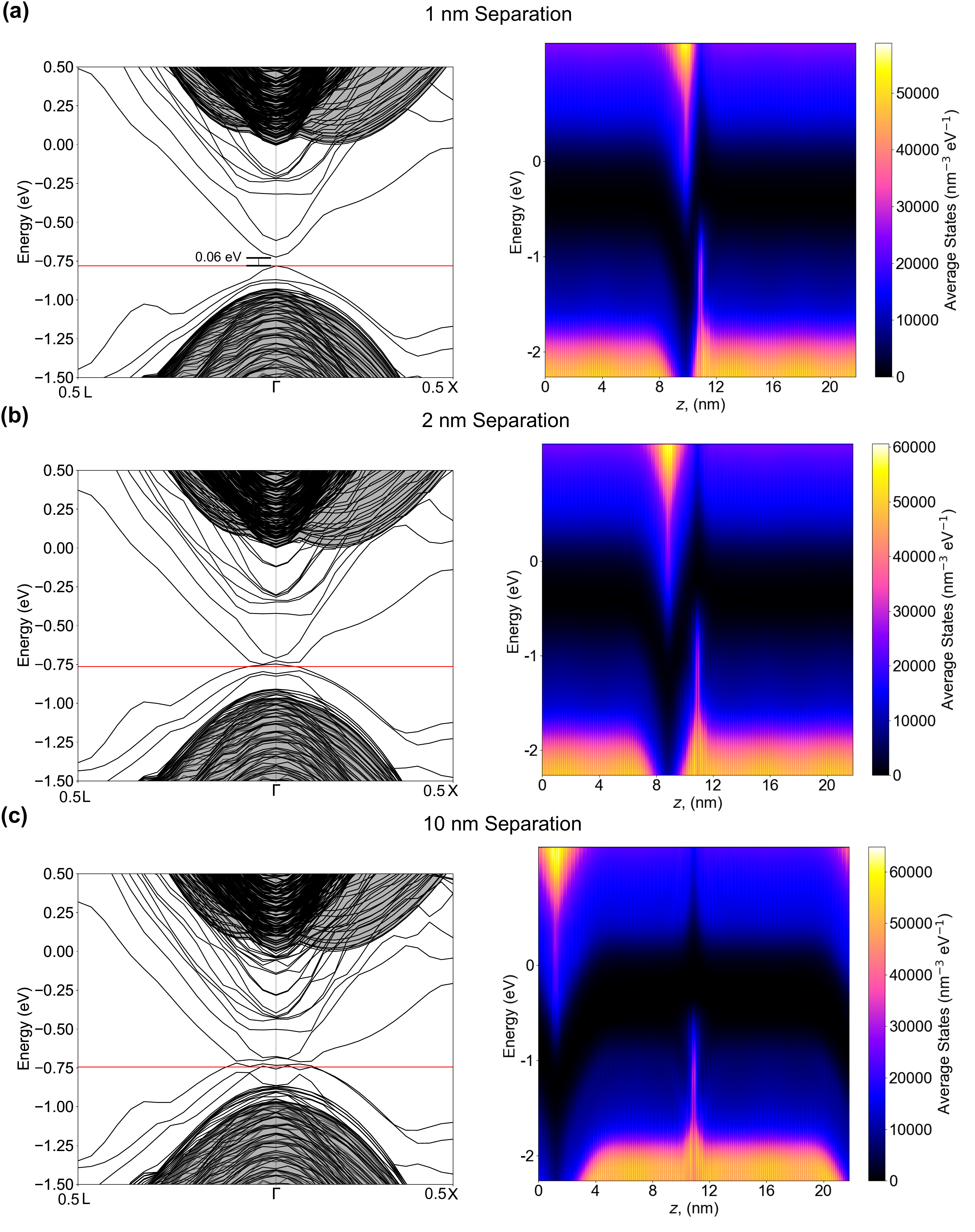}
\caption{The band structure and local density of states (LDOS) for aluminum and phosphorus $\delta$-doped layers separated by (a) 1 nm, (b) 2 nm and (c) 10 nm. The aluminum atoms have less suppression of the $\delta$ doped layer induced bands, resulting in overlap even at these lower separation distances. 
}
\label{fig:al-p-bandstruc-high} 
\end{figure*}
\subsection{Band structure and LDOS of aluminum-phosphorus $\delta$-doped layers}

In addition to the use of boron as an acceptor for $\delta$-doping a layer of silicon, aluminum has also been demonstrated as a potential dopant.\cite{radue2021alcl3} 
We examine how these aluminum-phosphorus structures may compare with boron-phosphorus structures in Fig.~\ref{fig:al-p-bandstruc-low} and \ref{fig:al-p-bandstruc-high}, looking at the band structure and the LDOS of the material. 
We see that in these systems the $\delta$-doped layer induced bands in the band structure are less suppressed by each other than the structures doped with phosphorus and boron, leading to more strongly overlapping band structures.
In appendix B, we explore the doping potentials that are generated by these boron-phosphorus and aluminum-phosphorus structures.
We demonstrate that boron induces more stress in the surrounding silicon than aluminum does.
This observation can be used to help explain the significantly lowered amount of suppression of the $\delta$-layer induced bands with aluminum compared to boron, and generally cleaner valence and conduction peaks in the resulting LDOS.
Stronger displacement from the boron $\delta$-doped layers can lead to small atomic variation in the position of nearby silicon atoms, which can help screen the potential from the $\delta$-doped layers from extending as far.  
Overall, however, these structures largely resemble the same behavior seen in the boron and phosphorus $\delta$-doped layer structures, with the energy between the acceptor and donor peaks lowering as the separation increases. 
The LDOS for these systems also remain similar, with distinct acceptor and donor peaks forming at $\delta$ layer separation distances $>$ 1 nm.
We hypothesize that an arsenic-boron system or arsenic-aluminum system (or additional combinations moving further down the periodic table) would continue the trend of lowered suppression of the $\delta$-layer induced bands. 

\begin{figure}
 \includegraphics[width=\columnwidth]{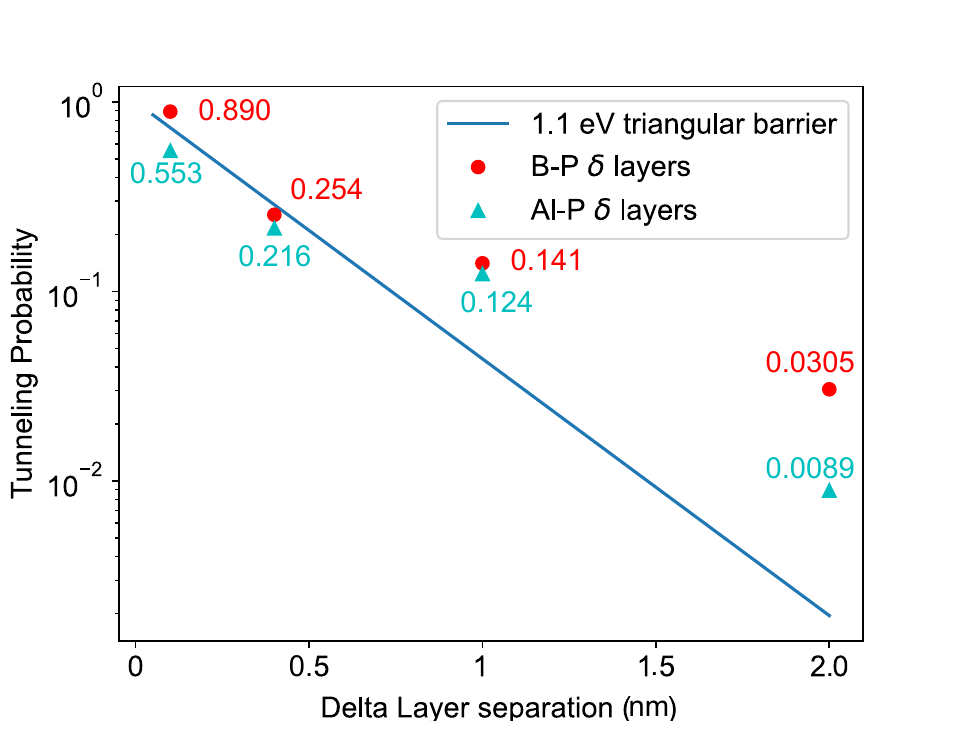}
\caption{Calculated tunneling probability for an electron in one of the $\delta$-layers tunneling into the other $\delta$-layer as a function of the separation between the $\delta$ layers. The tunneling probability calculated based on the actual potential measured from the DFT calculations is higher than a pure 1.1 eV triangular barrier of the same length due to the interaction between the $\delta$ layers.}
\label{fig:tunneling} 
\end{figure}
\subsection{Carrier transport through tunneling between $\delta$-doped layers}

Given the spatially separated nature of the $\delta$-layer peaks, if there is any charge carrier movement between the layers (absent the application of any bias on the system), one avenue is tunneling. 
The cost of an electron tunneling to the opposite polarity $\delta$-doped layer is dominated by overcoming the potential barrier induced by the interacting structure. 
We can approximate the tunneling probability of a single electron sitting with energy $E$ at the phosphorus $\delta$-doped layers moving to the boron $\delta$-doped layer using\cite{landau2013quantum}
\begin{equation}
  T(E) \simeq \exp \left[-2 \int_{z_1}^{z_2} |k(z)| dz \right],
\end{equation}
where $z_1$ and $z_2$ are the starting and ending point of the tunneling respectively (placed at the location of each of the $\delta$-doped layers), and 
\begin{equation}
  k(z) = \sqrt{\dfrac{2m^*}{\hbar^2}(V(z) - E)}.
\end{equation}
Here, $m^*$ is the effective mass of the electron, and $V(z)$ is the potential of the system as a function of the $z$, position.
We assume an effective mass of $m^* = 0.19 m_e$, where $m_e$ is the standard mass of an electron, matching the standard transverse mass of an electron in silicon, and the tunneling is solved for a single electron wavefunction  potential landscape $V(z)$.
We can conveniently abstract a $V(z)$ function from our DFT LDOS calculations (which are shown in Appendix C), and numerically integrate to calculate the tunneling probability in our systems. 
We then compare this to the tunneling that would be expected for a pure triangular 1.1 eV barrier in Fig.~\ref{fig:tunneling}.

Notably, due to the interaction between the two $\delta$-doped layers, the tunneling probability remains higher than would be expected for a pure triangular barrier.
This trend is particularly pronounced for longer separation distances.
We predict the tunneling probability of a boron and phosphorus $\delta$-doped layer separated by 1 and 2 nm is 0.141 and 0.0305, respectively.
While these are not large probabilities, they are significant enough that a reasonable number of electrons may tunnel through.
The tunneling rate in aluminum and phosphorus $\delta$-doped structures are lower than for boron and phosphorus structures due to the stronger peaks induced by less relaxation of nearby silicon.
This trend continues throughout the system, however, the slope of the Al-P $\delta$ layer tunneling remains above that of a 1.1 eV triangular barrier.
We note that these results are before any potential bias is applied to the system, which could be used to further manipulate tunneling rates. 
This tunneling probability calculation is based on an electron sitting at the bottom of the phosphorus $\delta$-layer potential well and assuming that it can tunnel to the top of the acceptor $\delta$ well.
This assumption of overlapping energy levels in the donor and acceptor is clearly true at high separation distances, but seems inaccurate for separations below 1 nm.
This is particularly true for boron $\delta$ layers, where the Fermi level does not sit between the valence and conduction bands, but seems to be caught within the B $\delta$ layer.
We therefore show these tunneling results merely to illustrate the possibility of an electron moving between layers, and compare how easy it would be in this system compared to a generic triangular potential well.  
The easiest electron movement within the given system would likely still be within a $\delta$ layer, making the behavior of the material essentially metallic around the delta layer. 
We next examine how easy it is to move electrons around the Fermi level of the system. 

\begin{figure}
 \includegraphics[width=\columnwidth]{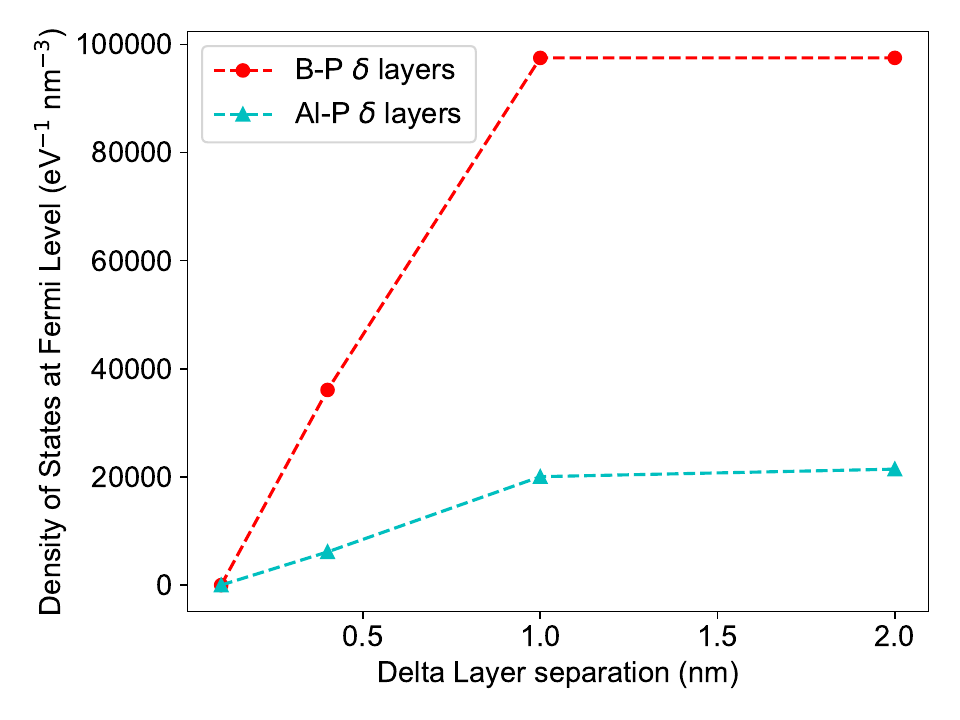}
\caption{The Density of States (DOS) at the Fermi level of the system as a function of separation between the $\delta$ layers. }
\label{fig:dos} 
\end{figure}

The density of states (DOS) at the Fermi level of the system gradually increases as the separation between the boron and phosphorus $\delta$-layers increases, as shown in Fig.~\ref{fig:dos}.
At lower separations, the DOS is quite low, as the $\delta$-layer induced bands are suppressed and the Fermi level is located within the band gap of the larger structure. 
As the separation between the delta layers increases, however, the $\delta$-layer induced valence and conduction bands increase in size and the Fermi level moves from the band gap into the valence bands. 
This causes the DOS at the Fermi level to increase with the separation between the materials until they appear to saturate at 1 nm separation. 
These higher separations are thus more likely to induce weak metallic-like behavior as electrons can easily move between bands in the valence band structure. 
While the aluminum-phosphorus structures have systematically lower density of states, the trends remain the same. 

\subsection{Considerations when comparing DFT results to experimental systems}
It should be noted that the DFT simulations provided in this work are necessarily an approximation of realistic systems with both acceptor and donor $\delta$-doped layers. 
Many of the errors are unfortunately inherent to any DFT-based approach for materials simulation of electronic properties. 
The band gap of the systems will be somewhat underestimated, an effect that has been well documented in various DFT calculations. 
We apply SCAN pseudopotentials\cite{sun2015strongly,yao2017plane} to combat this problem, but it has not been entirely solved, as evidenced by our prediction of a pure silicon supercell of 0.98 eV.
This may be particularly relevant at separation distances leading to us predicting a band structure that is continuous across the typical silicon band gap, such as the B-P 10 nm case (Fig.~\ref{fig:bandstruc_high_sep}c). 
There may, in fact, be slight band gaps in these materials that we cannot see without significantly more computationally expensive methods. 
Furthermore, there is some uncertainty in the exact placement of the Fermi level, which would be dependent on the specific doping of any given sample as well as the measurement temperature.  
When the Fermi level is placed firmly within the typical silicon band gap (as in the 0.1 and 0.4 nm separation cases for both B-P and Al-P layers), it is particularly ill-defined within DFT and should be taken more clearly as an indication that the valence band is fully occupied and the conduction band is fully unoccupied. 

There are three additional aspects of the results where the approximations necessary for DFT calculations become especially notable. 
The first is in the concentration effects of using a limited DFT supercell with periodic boundary conditions, which necessarily forces the concentration of dopants being modeled to be relatively high.
While this high concentration is laterally correct, i.e. roughly 1/4 monolayer coverage, it is unlikely to be correct perpendicular to the $\delta$ layer.
We are in fact modeling a silicon system with periodic $\delta$ layers every 21.8 nm rather than a single isolated pair of $\delta$ layers.
As recently explored by Campbell \textit{et al.}, \cite{campbell2023electronic} it can be difficult to fully disentangle the impact of $\delta$-doped layers from the high concentration of dopants being simulated.
Nonetheless, given the agreement between experimental characterization and DFT calculations of phosphorus $\delta$-doped layers,\cite{miwa2013direct,mazzola2014determining,mazzola2014disentangling,miwa2014valley,mazzola2018simultaneous,holt2020observation} it is reasonable to take these DFT results as good first-order approximations of the electronic structure. 

The second aspect of concern is unique to the current simulation of two interacting $\delta$-doped layers: the silicon region between the two $\delta$-doped layers.
Within our DFT simulations, we assume a pure, perfect crystalline silicon region between the two $\delta$-doped layers.
Due to the necessary low thermal budget processing to create such a structure, however, it is highly likely that this intermediate silicon region would not be perfectly crystalline, and may include a significant amount of defects. 
For example, silicon grown with APAM methods is known to contain oxygen and aluminum impurities.\cite{anderson2020JPhysMaterials}
This imperfect silicon structure and higher concentration of defects would likely reduce the interaction between the $\delta$-doped layers.
Further, our DFT calculations do not account for any dopant segregation that might occur during silicon epitaxy on top of the $\delta$-doped layers. 

The final aspect of concern is that while several of the systems show direct band gaps in the reported band diagrams, this represents an integration of the entire [100] direction within the system and the valence and conduction peaks are in fact spatially separated.
Since the simulated slab is so large in the [100] dimension, the corresponding Brillouin zone vector is small and sampling with more than one point would not lead to clear differences in the band diagram. 
This band diagram should be taken to represent the electronic structure of the entire system, and not correlated with one specific region of the model in real space. 
To understand how the electronic structure changes in real space, we present a light absorption thought experiment using the LDOS. 
Based on the LDOS of these systems, a photon shined onto the system would still have to overcome the intrinsic silicon band gap of $\approx$ 1 eV to be absorbed and generate a free electron and hole at any one location.  
Once generated, however, these charge carriers would be subject to the huge potential gradient between the $\delta$ layers and be swept to the nearest $\delta$ layer of the correct polarity. 
Similar analysis in InAs/GaSb superlattices, however, has shown that band gaps predicted with DFT of large supercells can match experimentally measured values, provided hybrid functionals or similar steps are taken to correct the band gap of the individual materials.\cite{garwood2017electronic,taghipour2018many} 
This gives us reason to believe our methodology provides useful reference points for future experimental characterization where these $\delta$ layers could be stacked to form a superlattice analogous to the well-established III-V superlattices, manipulating defects rather than band offsets. 

\section{Conclusion}

Motivated by recent progress in using APAM to create $\delta$-doped layers, we used DFT to predict the electronic structure of both interacting phosphorus and boron $\delta$-doped layers and interacting phosphorus and aluminum $\delta$-doped layers in silicon. 
At separations of less than $\approx$ 1 nm, we demonstrate that these $\delta$-doped layers will strongly suppress each other, leading to an effective band gap and electronic structure close to that of intrinsic silicon. 
As the separation between the layers increases to around 1 nm, the suppression decreases, and distinct $\delta$ potentials can be seen at the site of each layer, essentially creating a p-n diode with an intrinsic layer of silicon in the place of the depletion region. 
This decrease in suppression manifests as a change in the effective band gap that can be tailored by controlling the amount of silicon between the $\delta$-doped layers, ranging from 0.36 eV for 2 nm spacing to 0.86 eV for 0.1 nm spacing. 
However, the CBM and VBM are spatially separated, so carrier transport between $\delta$-doped layers likely requires tunneling or application of bias. 
Calculations of the $\delta$-doped layer potentials enable us to predict the tunneling rate between the two layers, finding that the tunneling at higher separations is greater than what would be expected for a pure 1.1 eV triangular barrier. 
These calculations thus provide a foundation for design of silicon electronics based on interacting $\delta$ layers.

Future work in this area would need to focus on experimentally creating both acceptor and donor $\delta$-doped layers stacked on top of each other in the same silicon device, which has not yet been demonstrated. 
The growth of clean, relatively defect free, intrinsic silicon between the two $\delta$-doped layers would be a particular challenge.
Further work is also needed investigating the electronic structure of $\delta$-doped layers in other group-IV materials such as SiGe, Ge, GeSn, etc. to gain a more complete understanding of the influence of varying strain and host material band structure.
Validation of the band structure could be undertaken with angle resolved photoemission spectroscopy (ARPES) as seen in previous experimental work looking at single $\delta$-doped layer silicon structures \cite{miwa2013direct,miwa2014valley,strand2025direct}.

\begin{acknowledgments}
We gratefully acknowledge useful conversations with 
Tzu-Ming Lu, Tommy Weingartner,
and Ezra Bussmann. 
This work was supported by the Laboratory Directed Research and Development Program at Sandia National Laboratories under project 226347.
This work was performed, in part, at the Center for Integrated Nanotechnologies, an Office of Science User Facility operated for the U.S. Department of Energy (DOE) Office of Science.
This article has been authored by an employee of National Technology \& Engineering Solutions of Sandia,
LLC under Contract No. DE-NA0003525 with the U.S. Department of Energy (DOE). The employee owns all right,
title and interest in and to the article and is solely responsible for its contents. The United States Government
retains and the publisher, by accepting the article for publication, acknowledges that the United States Government
retains a non-exclusive, paid-up, irrevocable, world-wide license to publish or reproduce the published form of this
article or allow others to do so, for United States Government purposes. The DOE will provide public access to
these results of federally sponsored research in accordance with the DOE Public Access Plan
https://www.energy.gov/downloads/doe-public-access-plan.
\end{acknowledgments}

\bibliography{references}

@article{Ward2020EDFA,
    author = {Ward, Daniel R. and Schmucker, Scott W. and Anderson, Evan M. and Bussmann, Ezra and Tracy, Lisa and Lu, Tzu-Ming and Maurer, Leon N. and Baczewski, Andrew and Campbell, Deanna M. and Marshall, Michael T. and Misra, Shashank},
    title = {Atomic Precision Advanced Manufacturing for Digital Electronics},
    journal = {EDFA Technical Articles},
    volume = {22},
    number = {1},
    pages = {4-10},
    year = {2020},
    month = {02},
    issn = {1537-0755},
    doi = {10.31399/asm.edfa.2020-1.p004},
    url = {https://doi.org/10.31399/asm.edfa.2020-1.p004},
    eprint = {https://dl.asminternational.org/edfa-tech/article-pdf/22/1/4/624359/edfa.2020-1.p004.pdf},
}

@article{carter2009electronic,
  title={Electronic structure models of phosphorus $\delta$-doped silicon},
  author={Carter, Damien J and Warschkow, Oliver and Marks, Nigel A and McKenzie, David R},
  journal={Physical Review B},
  volume={79},
  number={3},
  pages={033204},
  year={2009},
  publisher={APS}
}

@article{carter2011phosphorus,
  title={Phosphorus $\delta$-doped silicon: mixed-atom pseudopotentials and dopant disorder effects},
  author={Carter, Damien J and Marks, Nigel A and Warschkow, Oliver and McKenzie, David R},
  journal={Nanotechnology},
  volume={22},
  number={6},
  pages={065701},
  year={2011},
  publisher={IOP Publishing}
}

@article{drumm2013ab1,
  title={Ab initio calculation of valley splitting in monolayer $\delta$-doped phosphorus in silicon},
  author={Drumm, Daniel W and Budi, Akin and Per, Manolo C and Russo, Salvy P and Hollenberg, Lloyd CL},
  journal={Nanoscale research letters},
  volume={8},
  number={1},
  pages={1--11},
  year={2013},
  publisher={Springer}
}

@article{drumm2013ab2,
  title={Ab initio electronic properties of monolayer phosphorus nanowires in silicon},
  author={Drumm, DW and Smith, JS and Per, MC and Budi, A and Hollenberg, LCL and Russo, SP},
  journal={Physical review letters},
  volume={110},
  number={12},
  pages={126802},
  year={2013},
  publisher={APS}
}

@article{miwa2013direct,
  title={Direct measurement of the band structure of a buried two-dimensional electron gas},
  author={Miwa, Jill A and Hofmann, Philip and Simmons, Michelle Y and Wells, Justin W},
  journal={Physical review letters},
  volume={110},
  number={13},
  pages={136801},
  year={2013},
  publisher={APS}
}

@article{mazzola2014determining,
  title={Determining the electronic confinement of a subsurface metallic state},
  author={Mazzola, Federico and Edmonds, Mark T and H{\o}ydalsvik, Kristin and Carter, Damien John and Marks, Nigel A and Cowie, Bruce CC and Thomsen, Lars and Miwa, Jill and Simmons, Michelle Yvonne and Wells, Justin W},
  journal={ACS nano},
  volume={8},
  number={10},
  pages={10223--10228},
  year={2014},
  publisher={ACS Publications}
}

@article{mazzola2014disentangling,
  title={Disentangling phonon and impurity interactions in $\delta$-doped Si (001)},
  author={Mazzola, Federico and Polley, Craig M and Miwa, Jill A and Simmons, Michelle Y and Wells, Justin W},
  journal={Applied Physics Letters},
  volume={104},
  number={17},
  pages={173108},
  year={2014},
  publisher={American Institute of Physics}
}

@article{miwa2014valley,
  title={Valley splitting in a silicon quantum device platform},
  author={Miwa, Jill A and Warschkow, Oliver and Carter, Damien J and Marks, Nigel A and Mazzola, Federico and Simmons, Michelle Y and Wells, Justin W},
  journal={Nano letters},
  volume={14},
  number={3},
  pages={1515--1519},
  year={2014},
  publisher={ACS Publications}
}

@article{mazzola2018simultaneous,
  title={Simultaneous conduction and valence band quantization in ultrashallow high-density doping profiles in semiconductors},
  author={Mazzola, F and Wells, JW and Pakpour-Tabrizi, AC and Jackman, RB and Thiagarajan, B and Hofmann, Ph and Miwa, JA},
  journal={Physical review letters},
  volume={120},
  number={4},
  pages={046403},
  year={2018},
  publisher={APS}
}

@article{holt2020observation,
  title={Observation and origin of the $\Delta$ manifold in Si: P $\delta$ layers},
  author={Holt, Ann Julie and Mahatha, Sanjoy K and Stan, Raluca-Maria and Strand, Frode S and Nyborg, Thomas and Curcio, Davide and Schenk, Alex K and Cooil, Simon P and Bianchi, Marco and Wells, Justin W and others},
  journal={Physical Review B},
  volume={101},
  number={12},
  pages={121402},
  year={2020},
  publisher={APS}
}

@article{monkhorst1976special,
  title={Special points for Brillouin-zone integrations},
  author={Monkhorst, Hendrik J and Pack, James D},
  journal={Physical review B},
  volume={13},
  number={12},
  pages={5188},
  year={1976},
  publisher={APS}
}

@article{giannozzi2009quantum,
  title={QUANTUM ESPRESSO: a modular and open-source software project for quantum simulations of materials},
  author={Giannozzi, Paolo and Baroni, Stefano and Bonini, Nicola and Calandra, Matteo and Car, Roberto and Cavazzoni, Carlo and Ceresoli, Davide and Chiarotti, Guido L and Cococcioni, Matteo and Dabo, Ismaila and others},
  journal={Journal of physics: Condensed matter},
  volume={21},
  number={39},
  pages={395502},
  year={2009},
  publisher={IOP Publishing}
}

@article{radue2021alcl3,
  title={AlCl3-Dosed Si (100)-2$\times$ 1: Adsorbates, Chlorinated Al Chains, and Incorporated Al},
  author={Radue, Matthew S and Baek, Sungha and Farzaneh, Azadeh and Dwyer, KJ and Campbell, Quinn and Baczewski, Andrew D and Bussmann, Ezra and Wang, George T and Mo, Yifei and Misra, Shashank and others},
  journal={The Journal of Physical Chemistry C},
  year={2021},
  publisher={ACS Publications}
}

@article{dwyer2021area,
title={{B-Doped $\delta$-Layers and Nanowires from Area-Selective Deposition of BCl$_3$ on Si(100)}},
author = {Dwyer, Kevin J. and Baek, Sungha and Farzaneh, Azadeh and Dreyer, Michael and Williams, James R. and Butera, Robert E.},
journal={{ACS Applied Materials \& Interfaces}},
year = {2021},
doi = {10.1021/acsami.1c10616},
 
URL = {
https://doi.org/10.1021/acsami.1c10616
},
eprint = {
https://doi.org/10.1021/acsami.1c10616
}
}

@article{vskerevn2020bipolar,
  title={Bipolar device fabrication using a scanning tunnelling microscope},
  author={{\v{S}}kere{\v{n}}, Tom{\'a}{\v{s}} and K{\"o}ster, Sigrun A and Douhard, Bastien and Fleischmann, Claudia and Fuhrer, Andreas},
  journal={Nature Electronics},
  pages={1--7},
  year={2020},
  publisher={Nature Publishing Group}
}

@article{yao2017plane,
  title={Plane-wave pseudopotential implementation and performance of SCAN meta-GGA exchange-correlation functional for extended systems},
  author={Yao, Yi and Kanai, Yosuke},
  journal={The Journal of chemical physics},
  volume={146},
  number={22},
  pages={224105},
  year={2017},
  publisher={AIP Publishing LLC}
}

@article{sun2015strongly,
  title={Strongly constrained and appropriately normed semilocal density functional},
  author={Sun, Jianwei and Ruzsinszky, Adrienn and Perdew, John P},
  journal={Physical review letters},
  volume={115},
  number={3},
  pages={036402},
  year={2015},
  publisher={APS}
}

@article{campbell2022reaction,
  title={{Reaction pathways of BCl$_3 $ for acceptor delta-doping of silicon}},
  author={Campbell, Quinn and Dwyer, Kevin J and Baek, Sungha and Baczewski, Andrew D and Butera, Robert E and Misra, Shashank},
  journal={arXiv preprint arXiv:2201.11682},
  year={2022}
}

@article{campbell2023electronic,
    author = {Campbell, Quinn T. and Misra, Shashank and Baczewski, Andrew D.},
title = "{Electronic structure of boron and aluminum $\delta$-doped layers in silicon}",
journal = {Journal of Applied Physics},
volume = {134},
number = {4},
pages = {044401},
year = {2023},
month = {07},
issn = {0021-8979},
doi = {10.1063/5.0156832},
url = {https://doi.org/10.1063/5.0156832},
}

@article{bludau1974temperature,
  title={Temperature dependence of the band gap of silicon},
  author={Bludau, W and Onton, A and Heinke, W},
  journal={Journal of Applied Physics},
  volume={45},
  number={4},
  pages={1846--1848},
  year={1974},
  publisher={American Institute of Physics}
}

@inproceedings{lu2021path,
  title={Path Towards a Vertical TFET Enabled by Atomic Precision Advanced Manufacturing},
  author={Lu, Tzu-Ming and Gao, Xujiao and Anderson, Evan M and Mendez, Juan P and Campbell, DeAnna M and Ivie, Jeffrey A and Schmucker, Scott W and Grine, Albert and Lu, Ping and Tracy, Lisa A and others},
  booktitle={2021 Silicon Nanoelectronics Workshop (SNW)},
  pages={1--2},
  year={2021},
  organization={IEEE}
}

@inproceedings{gao2021modeling,
  title={Modeling and Assessment of Atomic Precision Advanced Manufacturing (APAM) Enabled Vertical Tunneling Field Effect Transistor},
  author={Gao, Xujiao and Mendez, Juan P and Lu, Tzu-Ming and Anderson, Evan M and Campbell, DeAnna M and Ivie, Jeffrey A and Schmucker, Scott W and Grine, Albert and Lu, Ping and Tracy, Lisa A and others},
  booktitle={2021 International Conference on Simulation of Semiconductor Processes and Devices (SISPAD)},
  pages={102--106},
  year={2021},
  organization={IEEE}
}

@article{katzenmeyer2020JMR, 
title={Assessing atomically thin delta-doping of silicon using mid-infrared ellipsometry}, 
volume={35}, 
DOI={10.1557/jmr.2020.155}, 
number={16}, 
journal={Journal of Materials Research}, 
publisher={Cambridge University Press}, 
author={Katzenmeyer, Aaron M. and Luk, Ting S. and Bussmann, Ezra and Young, Steve and Anderson, Evan M. and Marshall, Michael T. and Ohlhausen, James A. and Kotula, Paul and Lu, Ping and Campbell, DeAnna M. and et al.}, 
year={2020}, 
pages={2098–2105}
}

@article{anderson2020JPhysMaterials,
	doi = {10.1088/2515-7639/ab953b},
	url = {https://doi.org/10.1088/2515-7639/ab953b},
	year = 2020,
	month = {jun},
	publisher = {{IOP} Publishing},
	volume = {3},
	number = {3},
	pages = {035002},
	author = {Evan M Anderson and DeAnna M Campbell and Leon N Maurer and Andrew D Baczewski and Michael T Marshall and Tzu-Ming Lu and Ping Lu and Lisa A Tracy and Scott W Schmucker and Daniel R Ward and Shashank Misra},
	title = {Low thermal budget high-k/metal surface gate for buried donor-based devices},
	journal = {Journal of Physics: Materials}
	}

@article{halsey2022Robustness,
  author={Halsey, Connor and Depoy, Jessica and Campbell, DeAnna M. and Ward, Daniel R. and Anderson, Evan M. and Schmucker, Scott W. and Ivie, Jeffrey A. and Gao, Xujiao and Scrymgeour, David A. and Misra, Shashank},
  journal={IEEE Transactions on Device and Materials Reliability}, 
  title={Accelerated Lifetime Testing and Analysis of Delta-Doped Silicon Test Structures}, 
  year={2022},
  volume={22},
  number={2},
  pages={169-174},
  doi={10.1109/TDMR.2022.3152376}
  }

@INPROCEEDINGS{gao2020RTOperation,
  author={Gao, Xujiao and Tracy, Lisa A. and Anderson, Evan M. and Campbell, DeAnna M. and Ivie, Jeffrey A. and Lu, Tzu-Ming and Mamaluy, Denis and Schmucker, Scott W. and Misra, Shashank},
  booktitle={2020 International Conference on Simulation of Semiconductor Processes and Devices (SISPAD)}, 
  title={Modeling Assisted Room Temperature Operation of Atomic Precision Advanced Manufacturing Devices}, 
  year={2020},
  volume={},
  number={},
  pages={277-280},
  doi={10.23919/SISPAD49475.2020.9241642}
  }

@article{skeren2018nanotechnology,
	doi = {10.1088/1361-6528/aad7ab},
	url = {https://doi.org/10.1088/1361-6528/aad7ab},
	year = 2018,
	month = {aug},
	publisher = {{IOP} Publishing},
	volume = {29},
	number = {43},
	pages = {435302},
	author = {Tom{\'{a}}{\v{s}} {\v{S}}kere{\v{n}} and Nikola Pascher and Arnaud Garnier and Patrick Reynaud and Emmanuel Rolland and Aur{\'{e}}lie Thuaire and Daniel Widmer and Xavier Jehl and Andreas Fuhrer},
	title = {{CMOS} platform for atomic-scale device fabrication},
	journal = {Nanotechnology}
	}

@book{landau2013quantum,
  title={Quantum mechanics: non-relativistic theory},
  author={Landau, Lev Davidovich and Lifshitz, Evgenii Mikhailovich},
  volume={3},
  year={2013},
  publisher={Elsevier}
}

@article{young2023suppression,
	title={Suppression of Midinfrared Plasma Resonance Due to Quantum Confinement in $\delta$-Doped Silicon},
	author={Young, Steve M and Katzenmeyer, Aaron M and Anderson, Evan M and Luk, Ting S and Ivie, Jeffrey A and Schmucker, Scott W and Gao, Xujiao and Misra, Shashank},
	journal={Physical Review Applied},
	volume={20},
	number={2},
	pages={024043},
	year={2023},
	publisher={APS}
}

@article{garwood2017electronic,
	title={Electronic structure modeling of InAs/GaSb superlattices with hybrid density functional theory},
	author={Garwood, Tristan and Modine, Normand Arthur and Krishna, S},
	journal={Infrared Physics \& Technology},
	volume={81},
	pages={27--31},
	year={2017},
	publisher={Elsevier}
}

@article{taghipour2018many,
	title={Many-body perturbation theory study of type-II InAs/GaSb superlattices within the GW approximation},
	author={Taghipour, Zahra and Shojaee, Ezad and Krishna, Sanjay},
	journal={Journal of Physics: Condensed Matter},
	volume={30},
	number={32},
	pages={325701},
	year={2018},
	publisher={IOP Publishing}
}

@article{carter2013electronic,
	title={Electronic structure of two interacting phosphorus $\delta$-doped layers in silicon},
	author={Carter, DJ and Warschkow, O and Marks, NA and McKenzie, DR},
	journal={Physical Review B},
	volume={87},
	number={4},
	pages={045204},
	year={2013},
	publisher={APS}
}

@ARTICLE{wang2022natcomm,
       author = {{Wang}, Xiqiao and {Khatami}, Ehsan and {Fei}, Fan and {Wyrick}, Jonathan and {Namboodiri}, Pradeep and {Kashid}, Ranjit and {Rigosi}, Albert F. and {Bryant}, Garnett and {Silver}, Richard},
        title = "{Experimental realization of an extended Fermi-Hubbard model using a 2D lattice of dopant-based quantum dots}",
      journal = {Nature Communications},
         year = 2022,
        month = nov,
       volume = {13},
          eid = {6824},
        pages = {6824},
          doi = {10.1038/s41467-022-34220-w},
archivePrefix = {arXiv},
       eprint = {2110.08982},
 primaryClass = {quant-ph},
       adsurl = {https://ui.adsabs.harvard.edu/abs/2022NatCo..13.6824W}
}

@article{jones2023acsnano,
author = {Jones, Michael T. and Monir, Md Serajum and Krauth, Felix N. and Macha, Pascal and Hsueh, Yu-Ling and Worrall, Angus and Keizer, Joris G. and Kranz, Ludwik and Gorman, Samuel K. and Chung, Yousun and Rahman, Rajib and Simmons, Michelle Y.},
title = {Atomic Engineering of Molecular Qubits for High-Speed, High-Fidelity Single Qubit Gates},
journal = {ACS Nano},
volume = {17},
number = {22},
pages = {22601-22610},
year = {2023},
doi = {10.1021/acsnano.3c06668},
    note ={PMID: 37930801},

URL = { 
    
        https://doi.org/10.1021/acsnano.3c06668
    
    

},
eprint = { 
    
        https://doi.org/10.1021/acsnano.3c06668
    
    

}
}

@article{fuechsle2012single,
  title={A single-atom transistor},
  author={Fuechsle, Martin and Miwa, Jill A and Mahapatra, Suddhasatta and Ryu, Hoon and Lee, Sunhee and Warschkow, Oliver and Hollenberg, Lloyd CL and Klimeck, Gerhard and Simmons, Michelle Y},
  journal={Nature nanotechnology},
  volume={7},
  number={4},
  pages={242--246},
  year={2012},
  publisher={Nature Publishing Group UK London}
}

@article{fuechsle2010spectroscopy,
  title={Spectroscopy of few-electron single-crystal silicon quantum dots},
  author={Fuechsle, Martin and Mahapatra, S and Zwanenburg, Floris A and Friesen, Mark and Eriksson, MA and Simmons, Michelle Y},
  journal={Nature Nanotechnology},
  volume={5},
  number={7},
  pages={502--505},
  year={2010},
  publisher={Nature Publishing Group UK London}
}

@article{ruess2007realization,
  title={Realization of atomically controlled dopant devices in silicon},
  author={Rue{\ss}, Frank J and Pok, Wilson and Reusch, Thilo CG and Butcher, Matthew J and Goh, Kuan Eng J and Oberbeck, Lars and Scappucci, Giordano and Hamilton, Alex R and Simmons, Michelle Y},
  journal={Small},
  volume={3},
  number={4},
  pages={563--567},
  year={2007},
  publisher={Wiley Online Library}
}

@article{stock2020atomic,
  title={Atomic-scale patterning of arsenic in silicon by scanning tunneling microscopy},
  author={Stock, Taylor JZ and Warschkow, Oliver and Constantinou, Procopios C and Li, Juerong and Fearn, Sarah and Crane, Eleanor and Hofmann, Emily VS and Kolker, Alexander and McKenzie, David R and Schofield, Steven R and others},
  journal={ACS nano},
  volume={14},
  number={3},
  pages={3316--3327},
  year={2020},
  publisher={ACS Publications}
}

@article{campbell2021model,
  title={A model for atomic precision p-type doping with diborane on Si (100)-2$\times$ 1},
  author={Campbell, Quinn and Ivie, Jeffrey A and Bussmann, Ezra and Schmucker, Scott W and Baczewski, Andrew D and Misra, Shashank},
  journal={The Journal of Physical Chemistry C},
  volume={125},
  number={1},
  pages={481--488},
  year={2021},
  publisher={ACS Publications}
}

@article{campbell2022hole,
  title={Hole in one: Pathways to deterministic single-acceptor incorporation in Si (100)-2$\times$ 1},
  author={Campbell, Quinn and Baczewski, Andrew D and Butera, RE and Misra, Shashank},
  journal={AVS Quantum Science},
  volume={4},
  number={1},
  year={2022},
  publisher={AIP Publishing}
}

@article{roadmap,
doi = {10.1088/2399-1984/ada901},
url = {https://dx.doi.org/10.1088/2399-1984/ada901},
year = {2025},
month = {apr},
publisher = {IOP Publishing},
volume = {9},
number = {1},
pages = {012001},
author = {Schofield, Steven R and J Fisher, Andrew and Ginossar, Eran and Lyding, Joseph W and Silver, Richard and Fei, Fan and Namboodiri, Pradeep and Wyrick, Jonathan and Masteghin, Mateus G and Cox, David C and Murdin, Benedict N and Clowes, Steven K and G Keizer, Joris and Y Simmons, Michelle and Stemp, Holly G and Morello, Andrea and Voisin, Benoit and Rogge, Sven and A Wolkow, Robert and Livadaru, Lucian and Pitters, Jason and J Z Stock, Taylor and J Curson, Neil and Butera, Robert E and V Pavlova, Tatiana and Jakob, A M and Spemann, D and Räcke, P and Schmidt-Kaler, F and Jamieson, D N and Pratiush, Utkarsh and Duscher, Gerd and V Kalinin, Sergei and Kazazis, Dimitrios and Constantinou, Procopios and Aeppli, Gabriel and Ekinci, Yasin and Owen, James H G and Fowler, Emma and Moheimani, S O Reza and Randall, John and Misra, Shashank and A Ivie, Jeffrey and Allemang, Christopher R and Anderson, Evan M and Bussmann, Ezra and Campbell, Quinn and Gao, Xujiao and Lu, Tzu-Ming and Schmucker, Scott W},
title = {Roadmap on atomic-scale semiconductor devices},
journal = {Nano Futures},
}

@article{anderson2025integration,
    author = {Anderson, E. M. and Allemang, C. R. and Leenheer, A. J. and Schmucker, S. W. and Ivie, J. A. and Campbell, D. M. and Lepkowski, W. and Gao, X. and Lu, P. and Arose, C. and Lu, T.-M. and Halsey, C. and England, T. D. and Ward, D. R. and Scrymgeour, D. A. and Misra, S.},
    title = {Direct integration of atomic precision advanced manufacturing into middle-of-line silicon fabrication},
    journal = {Applied Physics Reviews},
    volume = {12},
    number = {4},
    pages = {041402},
    year = {2025},
    month = {10},
    issn = {1931-9401},
    doi = {10.1063/5.0278639},
    url = {https://doi.org/10.1063/5.0278639}
}

@article{mendez2025exploring,
  title={Exploring transport mechanisms in atomic precision advanced manufacturing enabled pn junctions},
  author={Mendez, Juan P and Gao, Xujiao and Ivie, Jeffrey A and Owen, James HG and Kirk, Wiley P and Randall, John N and Misra, Shashank},
  journal={Journal of Applied Physics},
  volume={137},
  number={13},
  year={2025},
  publisher={AIP Publishing}
}

@article{strand2025direct,
  title={Direct Observation of 2DEG States in Shallow Si: Sb $\delta$-Layers},
  author={Strand, Frode S and Cooil, Simon P and Campbell, Quinn T and Flounders, John J and R{\o}st, H{\aa}kon I and {\AA}sland, Anna Cecilie and Skarpeid, Alv Johan and Stalsberg, Marte P and Hu, Jinbang and Bakkelund, Johannes and others},
  journal={The Journal of Physical Chemistry C},
  volume={129},
  number={2},
  pages={1339--1347},
  year={2025},
  publisher={ACS Publications}
}

\clearpage
\widetext
\appendix

\section{Band Structures of isolated single $\delta$-doped layers}

In this appendix, we predict the band structure of a single $\delta$-doped layer, with either boron or phosphorus dopants, using SCAN pseudopotentials for higher accuracy band gaps.
We use the same supercell and calculation parameters as described in Sec.~\ref{sec:methods} of the main text. 
This allows for easy comparison with our results in Sec.~\ref{sec:results} to determine what is an impact of interaction between the two $\delta$-doped layers.
Our results are in many ways similar to previous DFT of phophorus $\delta$-doped layers \cite{carter2009electronic,carter2011phosphorus,drumm2013ab1,drumm2013ab2} and boron $\delta$-doped layers \cite{campbell2023electronic}, but these results are the first to use SCAN pseudopotentials \cite{sun2015strongly}, which provide more accurate band gaps for semiconductors. 
\begin{figure*}
 \includegraphics[width=\textwidth]{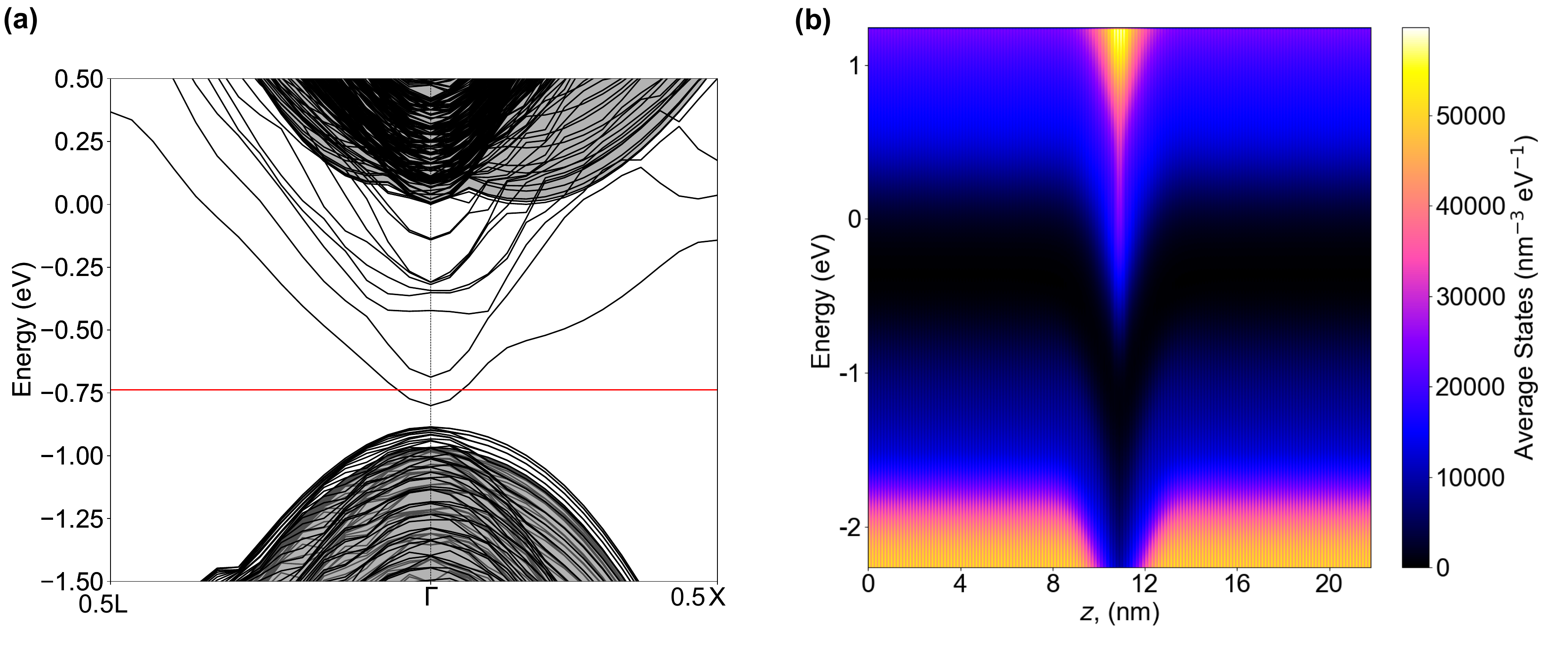}
\caption{The (a) band structure and (b) local density of states for a single phoshporus $\delta$-doped layer, using the same supercell and pseudopotentials as the main text.
}
\label{fig:p-delta} 
\end{figure*}

\begin{figure*}
 \includegraphics[width=\textwidth]{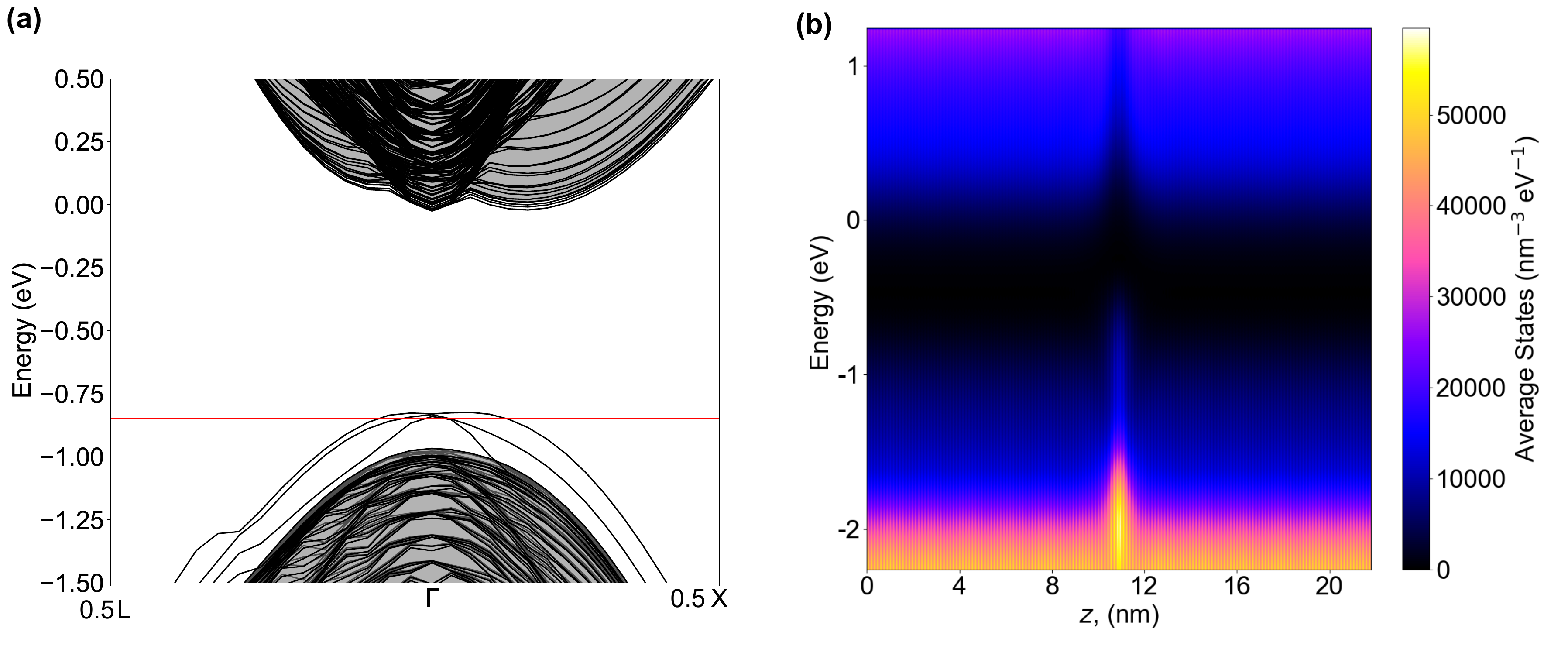}
\caption{The (a) band structure and (b) local density of states for a single boron $\delta$-doped layer, using the same supercell and pseudopotentials as the main text.
}
\label{fig:b-delta} 
\end{figure*}

\begin{figure*}
 \includegraphics[width=\textwidth]{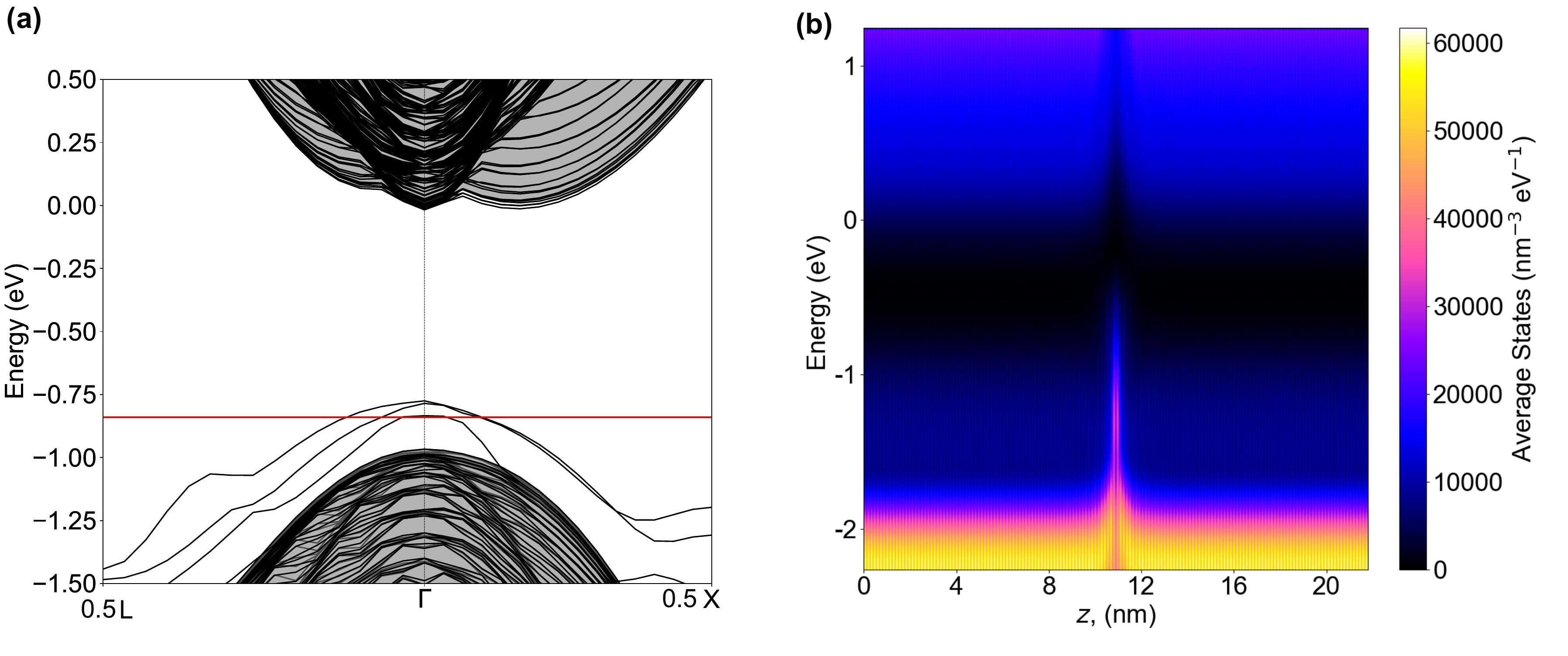}
\caption{The (a) band structure and (b) local density of states for a single aluminum $\delta$-doped layer, using the same supercell and pseudopotentials as the main text.
}
\label{fig:al-delta} 
\end{figure*}

The band structure and LDOS of a phosphorus $\delta$-doped layer is shown in Fig.~\ref{fig:p-delta}, a boron $\delta$-doped layer band structure is shown in Fig.~\ref{fig:b-delta}, and an aluminum $\delta$-doped layer band structure is shown in Fig.~\ref{fig:al-delta}.
For each structure, the effects are limited entirely to the conduction band (in the case of phosphorus) or the valence band (in the case of boron).
Naively, we might expect that the effects of including both a phosphorus and boron $\delta$-doped layer would be a superposition of the phosphorus conduction band and the boron conduction bands. 
This description only resembles the band structures when the $\delta$-doped layers are separated by $\geq$ 2.0 nm, indicating that interaction between the layers is indeed causing significant suppression of the bands at lower separations.
It is notable that SCAN pseudopotentials predict that the $\delta$ potential induced by the phosphorus $\delta$-doped layer is much more significant in magnitude than previous predictions.
While the exact magnitude is likely an overestimate, it does indicate that phosphorus $\delta$-doped layers should have an extremely large profile within our calculations. 
This indicates that the suppression of the phosphorus $\delta$-potential at low separation distances is even more significant than it may appear at first glance.

\section{Doping potentials for the boron and aluminum $\delta$-doped layers interacting with a phosphorus $\delta$-doped layers}

\begin{figure*}
 \includegraphics[width=\textwidth]{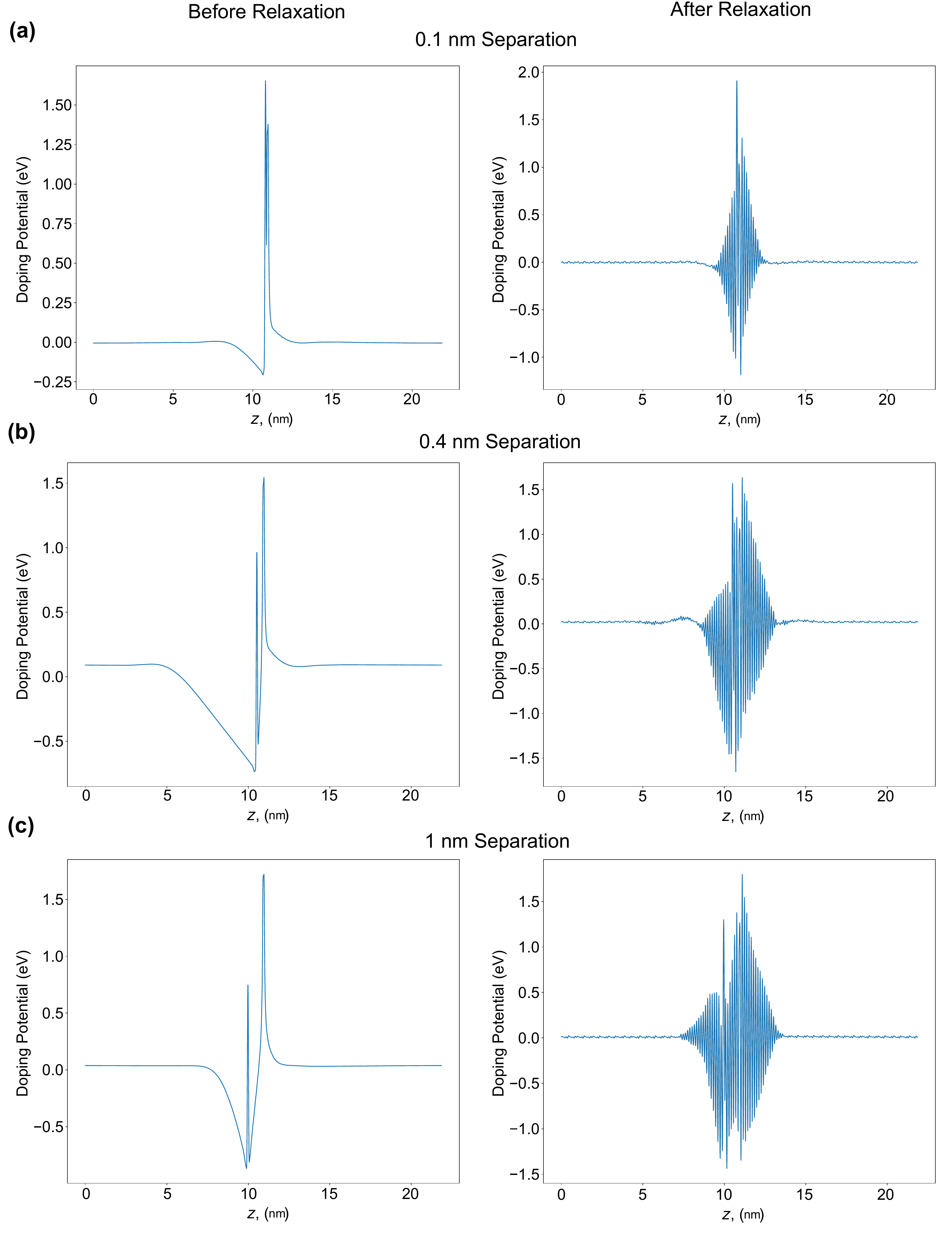}
\caption{The doping potential of structures with (a) 0.1 nm, (b) 0.4 nm, (c) 1.0 nm~separation between the boron and phosphorus $\delta$-doped layers.
The left column displays the potential before the atomic structures are allowed to relax to respond to stress, and the right column is the potential after the structures are allowed to relax.
}
\label{fig:doping_pot_low_sep} 
\end{figure*}

\begin{figure*}
 \includegraphics[width=\textwidth]{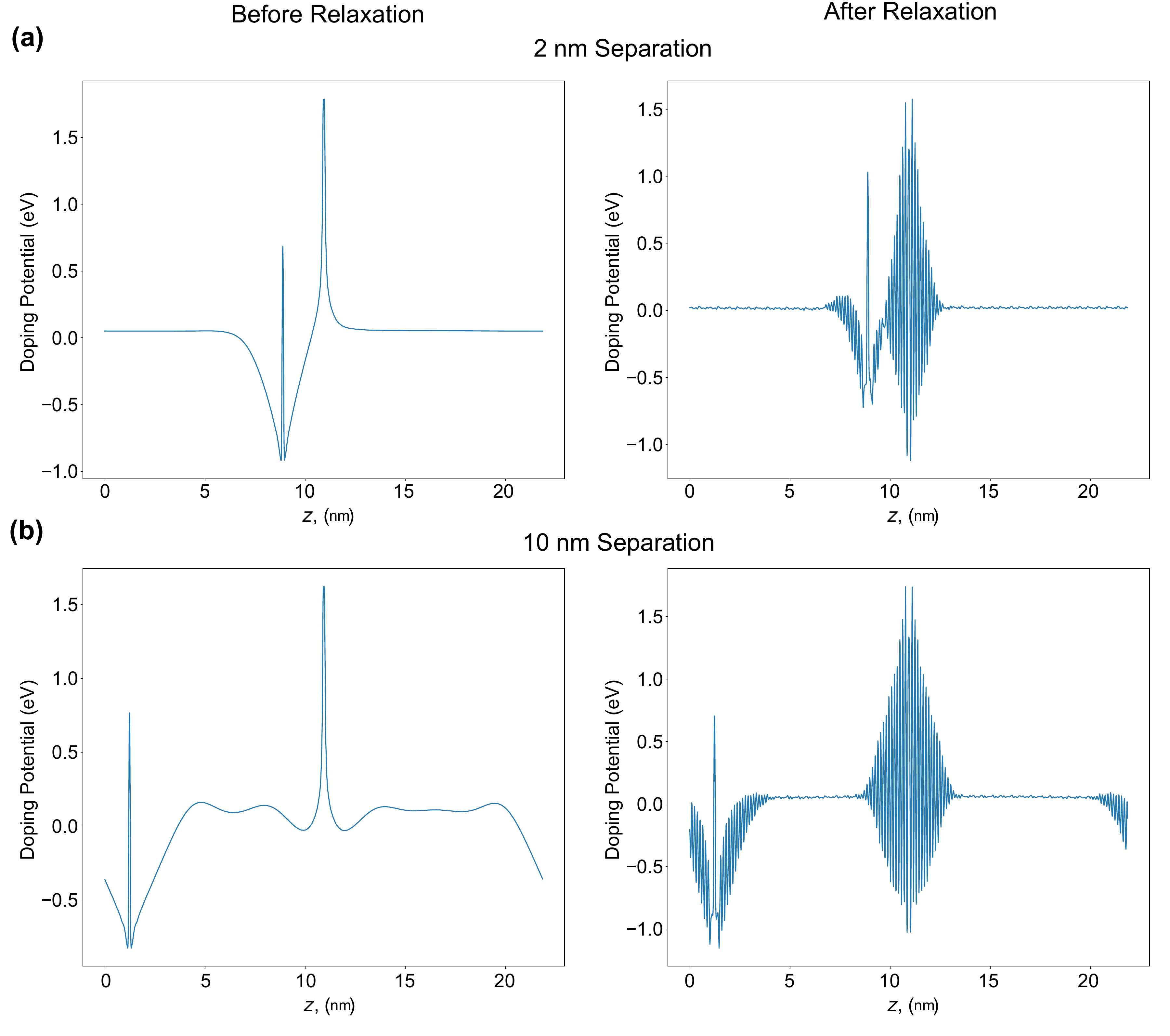}
\caption{The doping potential of structures with (a) 2.0 nm, (b) 10.0 nm~separation between the boron and phosphorus $\delta$-doped layers.
The left column displays the potential before the atomic structures are allowed to relax to respond to stress, and the right column is the potential after the structures are allowed to relax.
}
\label{fig:doping_pot_high_sep} 
\end{figure*}

To further elucidate the nature of the $\delta$-doped layers interaction, we plot the doping potentials for each separation distance in Figs.~\ref{fig:doping_pot_low_sep} and ~\ref{fig:doping_pot_high_sep}.
These potentials are calculated by subtracting the planar averaged electrostatic potential of a pure silicon supercell from the planar averaged electrostatic potential of the system with boron and phosphorus $\delta$-doped layers.
The doping potentials calculated in the left column of each figure represent the potential before the system is allowed to relax in response to the stress that the doping atoms induce in the system. 
These graphs more or less follow the pattern that would be expected from the LDOS calculated in Figs.~\ref{fig:bandstruc_low_sep} and \ref{fig:bandstruc_high_sep}.
In the right column, however, we display the potential after the system is allowed to relax in response to stress induced by the doping atoms. 
These potentials have a much more oscillatory character around the $\delta$-doped layers, which is a direct result of atomic movement in response to the strain slightly altering the periodicity of the nearby silicon atoms. 
Similar to what is seen in Ref. \onlinecite{campbell2023electronic}, boron $\delta$-doped layers feature particularly strong oscillation with the resulting $\delta$ potential being less well defined within the material compared to the phosphorus $\delta$-doped layers.
This is especially clear in the well separated structures at 2.0 and 10.0 nm~separation shown in Fig.~\ref{fig:doping_pot_high_sep}.

\begin{figure*}
 \includegraphics[width=\textwidth]{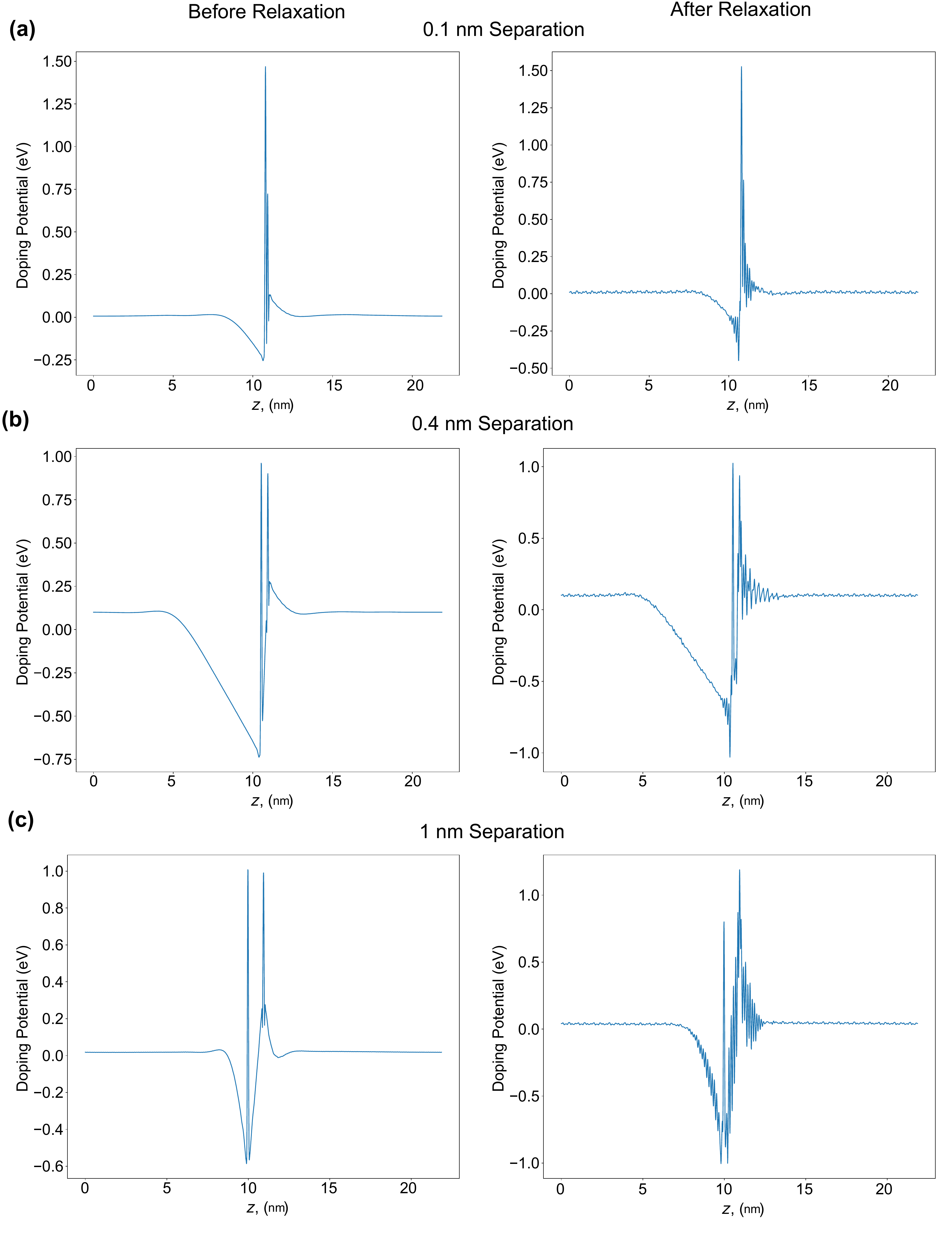}
\caption{The doping potential of Al-P structures with (a) 0.1 nm, (b) 0.4 nm, and (c) 1.0 nm~separation between the boron and phosphorus $\delta$-doped layers.
The left column displays the potential before the atomic structures are allowed to relax to respond to stress, and the right column is the potential after the structures are allowed to relax.
}
\label{fig:al_doping_pot} 
\end{figure*}

\begin{figure*}
 \includegraphics[width=\textwidth]{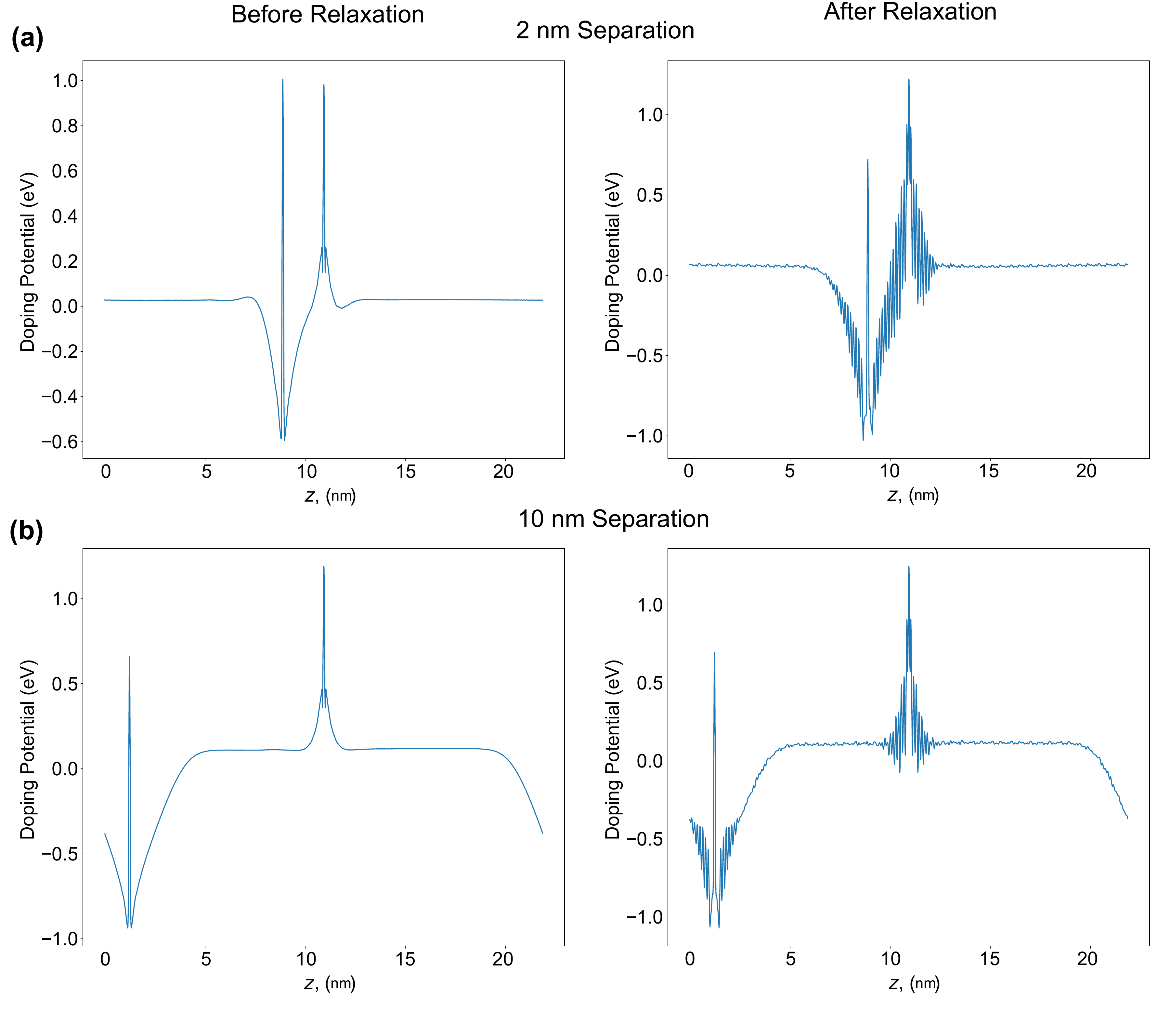}
\caption{The doping potential of Al-P structures with (a) 2.0 nm, (b) 10.0 nm~separation between the boron and phosphorus $\delta$-doped layers.
The left column displays the potential before the atomic structures are allowed to relax to respond to stress, and the right column is the potential after the structures are allowed to relax.
}
\label{fig:al_doping_pot_high_sep} 
\end{figure*}

We finally examine the doping potentials of an aluminum $\delta$-doped layer interacting with a phosphorus $\delta$-doped layer, as shown in Fig.~\ref{fig:al_doping_pot} and \ref{fig:al_doping_pot_high_sep}.
The oscillatory nature of the doping potentials of the relaxed structures is much reduced, indicating much less stress is induced in response to the aluminum $\delta$-doped layer.

To provide a quantitative comparison between the two systems, we define a decay width $\sigma_{doping}$, 
\begin{equation}
    \sigma_{doping} = z_{thr} - z_{B/Al} ,
\end{equation}
where $z_{B/Al}$ is the z position where the B or Al $\delta$ layer is centered, and $z_{thr}$ is the threshold z position, at which the potential oscillation is $<$ than a specified value. 
We choose the threshold to be a z value with potential oscillations $<$ 0.02 V.
We find that for B $\delta$ layers the average $\sigma_{doping} = 2.2$ nm, whereas for Al $\delta$ layers, the average $\sigma_{doping} = 1.6$ nm.
This indicates that B $\delta$ layers induce atomic displacement in the surrounding silicon an average of 0.6 nm further away than Al layers do, which may serve as a partial explanation for why the Al-P structures tend to have more defined peaks in the LDOS than B-P structures do. 

\section{Tunneling potentials for the boron and aluminum $\delta$-doped layers interacting with a phosphorus $\delta$-doped layers}

\begin{figure*}
 \includegraphics[width=\textwidth]{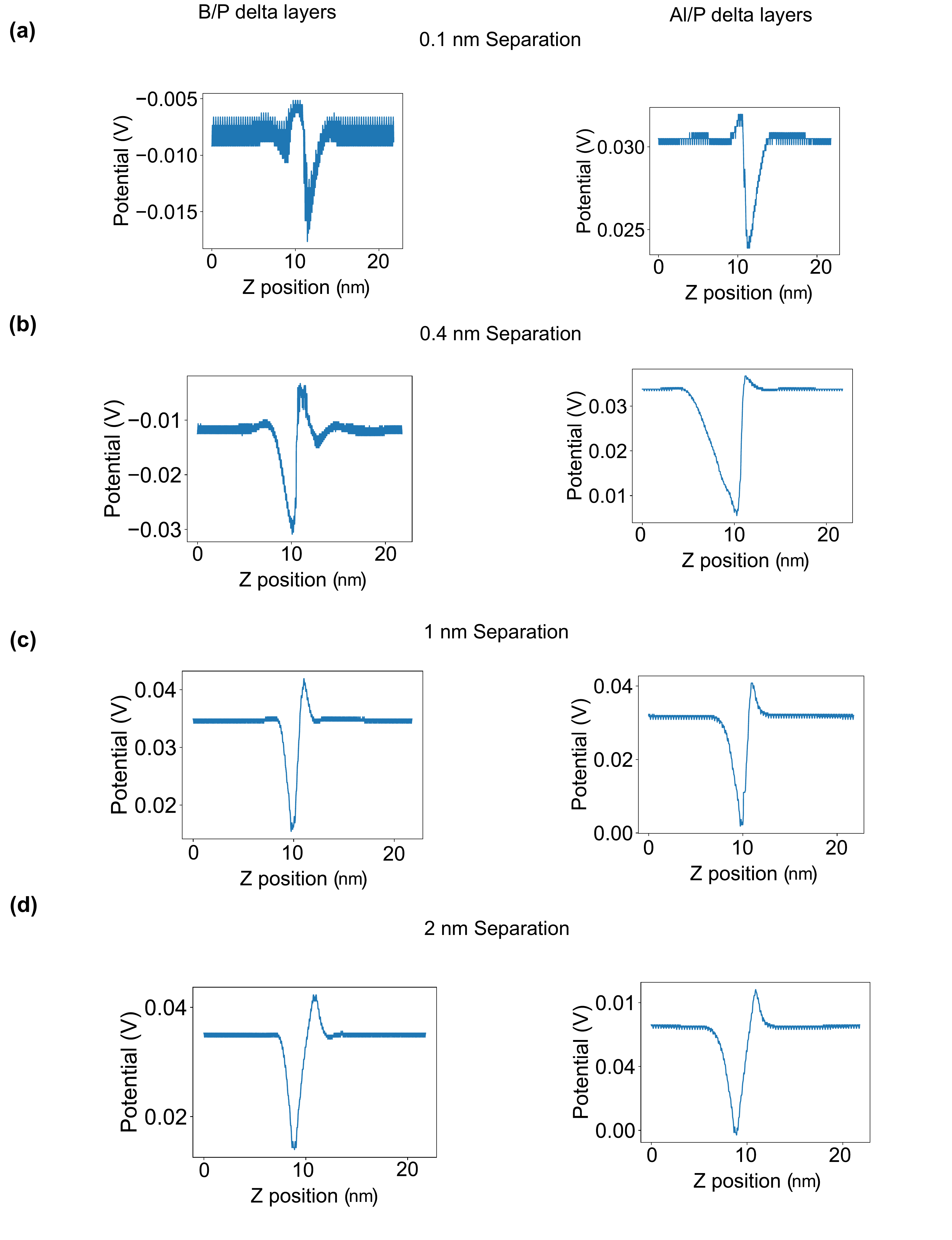}
\caption{The conduction band potential of B-P and Al-P structures with (a) 0.1 nm, (b) 0.4 nm, (c) 1.0 nm, (d) 2.0 nm~separation between the boron and phosphorus $\delta$-doped layers.
These calculations are all for relaxed structures. 
}
\label{fig:tunnel_pot} 
\end{figure*}

From the LDOS diagrams in Figures 2-5, we can extract the outline of the conduction band minima (and corresponding valence band  maxima with a potential shift) looks like. 
We then use this as an input potential for the tunneling calculations described in Sec.~\ref{sec:results}C. 
We show the extracted $V(z)$ plots for each system used in our tunneling calculations in Fig.~\ref{fig:tunnel_pot}.

\end{document}